\documentclass[nofootinbib,twocolumn]{revtex4-1}
\usepackage{graphicx,natbib}
\usepackage{url}
\usepackage{color}
\usepackage{rotating}
\usepackage{amssymb,amsmath}

\newcommand{\apjs}{ApJS}
\newcommand{\mnras}{MNRAS}

\newcommand{\Mpc}{{\rm Mpc}}
\newcommand{\Gpc}{{\rm Gpc}}
\newcommand{\degr}{\hbox{$^\circ$}}
\newcommand{\lcdm}{$\Lambda$CDM\;}

\newcommand{\DRsup}{\Delta_{R, {\rm sup}}^2(k)}
\newcommand{\DRunsup}{\Delta_{R, {\rm unsup}}^2(k)}

\newcommand{\dccl}{\Delta \chi^2_{C_{\ell}}}
\newcommand{\dcs}{\Delta \chi^2_{S_{1/2}}}
\newcommand{\pcl}{P(C_{\ell})}
\newcommand{\pss}{P(S_{1/2})}
\newcommand{\logkc}{\log_{10}(k_c/(h/\Mpc))}

\newcommand{\alm}{a_{\ell m}}
\newcommand{\Cl}{C_\ell}

\newcommand{\fsky}{f_{\rm sky}}

\begin{document}

\title{Detectability of large-scale power suppression in the galaxy distribution}

\author{Cameron Gibelyou}
\affiliation{Department of Physics, University of Michigan, 
450 Church St, Ann Arbor, MI 48109-1040}

\author{Dragan Huterer}
\affiliation{Department of Physics, University of Michigan, 
450 Church St, Ann Arbor, MI 48109-1040}

\author{Wenjuan Fang}
\affiliation{Department of Physics, University of Michigan, 
450 Church St, Ann Arbor, MI 48109-1040}

\date{\today}

\begin{abstract}
  Suppression in primordial power on the Universe's largest observable
  scales has been invoked as a possible explanation for large-angle
  observations in the cosmic microwave background, and is allowed or
  predicted by some inflationary models. Here we investigate the
  extent to which such a suppression could be confirmed by the
  upcoming large-volume redshift surveys. For definiteness, we study a
  simple parametric model of suppression that improves the fit of the
  vanilla \lcdm model to the angular correlation function measured by
  WMAP in cut-sky maps, and at the same time improves the fit to the
  angular power spectrum inferred from the maximum-likelihood analysis
  presented by the WMAP team. We find that the missing power at large
  scales, favored by WMAP observations within the context of this
  model, will be difficult but not impossible to rule out with a
  galaxy redshift survey with large volume ($\sim 100~\Gpc^3$).  A key
  requirement for success in ruling out power suppression will be
  having redshifts of most galaxies detected in the imaging survey.
\end{abstract}

\maketitle

\section{Introduction}\label{sec:intro}

Measurements of the angular power spectrum of the cosmic microwave
background (CMB) anisotropies from the WMAP experiment have been used
to constrain the standard cosmological parameters to unprecedented
accuracy \cite{Spergel2003,Spergel2006,wmap5,wmap7}.  At the same
time, several anomalies have been observed, one of which is the
missing power above 60 degrees on the sky in the maps where the
galactic plane has been masked
\cite{Spergel2003,wmap123,wmap12345}. This is unexpected not only
because skies with such lack of large-scale power are expected with
the probability of about 0.03\% in the standard Gaussian, isotropic
model \cite{wmap123,wmap12345,Sarkar}, but for two other
reasons. First, the missing power occurs on the largest observable
scales, where a cosmological origin is arguably most likely. Second,
missing correlations are inferred from cut-sky (i.e.\ masked) maps of
the CMB, which makes the results insensitive to assumptions about what
lies behind the cut.
For review of the missing correlations (and other so-called ``large-angle
anomalies'' in the CMB), see \cite{CHSS_review}; for debate on this issue, see
\cite{CHSS_review,Efstathiou2009,Pontzen_Peiris,Aurich_Lustig}; for signatures
of the anomalies in future polarization observations, see \cite{Dvorkin}.

In this paper we study the possibility that the {\it primordial} power
spectrum is suppressed at large scales. This explanation has been
invoked before in order to explain the low power in the multipole
spectrum (e.g.\ \cite{Contaldi}). In the meantime, observations have
made it apparent that the harmonic-space quadrupole and octopole are
only moderately low (e.g.~\cite{O'Dwyer2004,Hinshaw:2006ia}), and it
is really a range of low multipoles that conspire to produce the
vanishing $C(\theta)$. Specifically, as discussed in \cite{wmap12345},
there is a cancellation between the combined contributions of
$C_2$,...,$C_5$ and the contributions of $\Cl$ with $\ell\geq 6$.  It
is this conspiracy that is most disturbing, since it violates the
independence of the $\Cl$ of different $\ell$ that defines statistical
isotropy.

Note however that it is {\it a priori} not at all clear that
suppression in the large-scale power can explain the WMAP observations
on large scales. While the missing large-angle correlations in the
angular two-point correlation function of the CMB $C(\theta)$ could be
trivially explained by the missing primordial power, a large
suppression would lower the harmonic power spectrum $\Cl$, inferred
using the maximum-likelihood estimator, too much to be consistent with
observations. [We discuss this in Sections \ref{sec:statistical} and
\ref{sec:constraints} below.]

In this paper we perform a two-pronged analysis. First, we adopt a simple
parametric model for the suppression, and perform a detailed analysis to find
the suppressed power spectrum that improves the fit of the vanilla \lcdm model
to the angular correlation function measured by WMAP in cut-sky maps, and at
the same time improves the fit to the angular power spectrum inferred from the
maximum-likelihood analysis presented by the WMAP team. Second, we address the
following question: {\it if} the CMB observations are telling us that the
three-dimensional primordial power spectrum is indeed suppressed at large
scales (and our adopted model for the suppression is at work), could this
effect be confirmed in redshift surveys, with observations of suppressed
clustering of galaxies on the largest scales? 

It is important to note that we do not concern ourselves with
questions recently discussed in the literature as to whether the
full-sky or the cut-sky measurements are more robust. It could be the
case that one of these measurements, full-sky or cut-sky, is correct
while the other is not due to some type of systematic error; it could
also be that both of these measurements are correct (in which case the
assumption of statistical isotropy is arguably on less firm
footing). We consider these possibilities separately in order to get a
rough idea on what scale the data favor power suppression in either
case. Regardless of which of these possibilities is true, however, our
results regarding the detectability of power suppresion, presented in
Sec.~\ref{sec:future}, are valid.

The paper is organized as follows. In Sec.~\ref{sec:prelim} we do a
preliminary investigation in which we attempt to reconstruct the suppression
of the primordial power spectrum (and, correspondingly, the matter power
spectrum) directly from CMB angular power spectrum measurements (the
$C_{\ell}$). In Sec.~\ref{sec:suppressed}, we take the complementary approach,
parameterizing the suppression and finding how it affects the $C_{\ell}$ and
$C(\theta)$. Sec.~\ref{sec:statistical} quantifies how well a given suppressed
model fits CMB data in both $C_{\ell}$ and
$C(\theta)$. Sec.~\ref{sec:constraints} provides a discussion of the results
that we obtain from this analysis; we find that large-scale suppression of
power can significantly increase the likelihood of the observed CMB at large
scales. Finally, in Sec.~\ref{sec:future}, we discuss the possibility of
detecting suppression in the matter power spectrum with an upcoming
large-volume redshift survey. We conclude in Sec.~\ref{sec:conclusions}.

\section{Suppressed power: preliminary investigations}\label{sec:prelim}

Let us first review the basic way in which the primordial power
spectrum determines fluctuations in the CMB observed today.  The CMB
temperature anisotropies are decomposed into spherical harmonics with
coefficients $a_{\ell m}$
\begin{equation}
\frac{\Delta T}{T}(\theta,\phi) = \sum_{\ell=2}^\infty \sum_{m=-\ell}^\ell a_{\ell m} Y_{\ell m}(\theta, \phi),
\label{eqn:SHdecomposition}
\end{equation}
where $T$ is the average temperature of the CMB. The angular power
spectrum, which quantifies the contribution to the variance of the
temperature fluctuations at each $\ell$, is then given by the
coefficients $\Cl$ where, assuming statistical isotropy,
$\left \langle \alm a_{\ell' m'} \right \rangle=\Cl
\delta_{\ell\ell'}\delta_{mm'}$.
We will also consider the angular two-point correlation function
\begin{equation}
C(\theta)\equiv \left\langle \frac{\Delta T}{T}(\hat n)
\frac{\Delta T}{T}(\hat n')\right\rangle_{\hat n\cdot\hat n'=\cos \theta},
\label{eq:ctheta}
\end{equation}
where we have assumed statistical isotropy, and the expectation is taken over
the ensemble of universes. $C(\theta)$ is related
to the anisotropy power spectrum by
\begin{equation}
 C(\theta) = \frac{1}{4\pi}\sum_{\ell=2}^\infty (2\ell +1) C_\ell P_\ell (\cos
 \theta).
\label{eq:Ctheta_vs_Cl}
\end{equation}

The angular power spectrum $\Cl$ is directly related to the primordial
power spectrum of curvature perturbations laid down by inflation. The
$C_{\ell}$ are given in terms of the primordial power spectrum by
\begin{equation}
\frac{\ell (\ell + 1) C_{\ell}}{2 \pi} = \int d(\ln k) 
\left \lbrack T_{\ell}(k) \right \rbrack ^2 \Delta_{R}^2 (k),
\label{eqn:ClDelta}
\end{equation}
where $T_{\ell}(k)$ is the transfer function and $\Delta_R^2(k)$ is
the dimensionless curvature power spectrum
\begin{equation}
\Delta_{R}^2 (k) \equiv \frac{k^3 P_{R}(k)}{2 \pi^2},
\label{eqn:Delta2def}
\end{equation}
where $P_R(k)$ is the curvature power spectrum which, at late times
and on subhorizon scales, is related to the matter density power
spectrum $P(k)$ via $P_R(k) \propto k^{-4} P(k)$.

We would like to infer the primordial curvature power spectrum
$\Delta^2_R(k)$ given the angular power spectrum $C_\ell$ measured
from the CMB. There are two approaches we could take to dealing with
Eq.~(\ref{eqn:ClDelta}): the inverse problem (discussed in this
section) and the parametric forward problem of starting with various
power spectra and attempting to fit the $C_\ell$ (discussed in the
next section and pursued in the rest of the paper).

The first option is to directly calculate $\Delta^2_R(k)$ from the
measured angular power spectrum $C_{\ell}$. This inverse problem,
where we know the result of the integration but not the integrand, is
difficult because the primordial power spectrum $\Delta_R^2(k)$ is a
three-dimensional quantity while the CMB angular power spectrum
$C_{\ell}$ is a two-dimensional, projected quantity. When the problem
is discretized, as described below, it becomes clear that the problem
is underdetermined and ill-conditioned, as is typical for inverse
problems: small changes in the observed $C_\ell$ typically lead to
large changes in the inferred $\Delta^2_R(k)$.

\begin{figure}[t]
\begin{center}
\includegraphics[width=.45\textwidth]{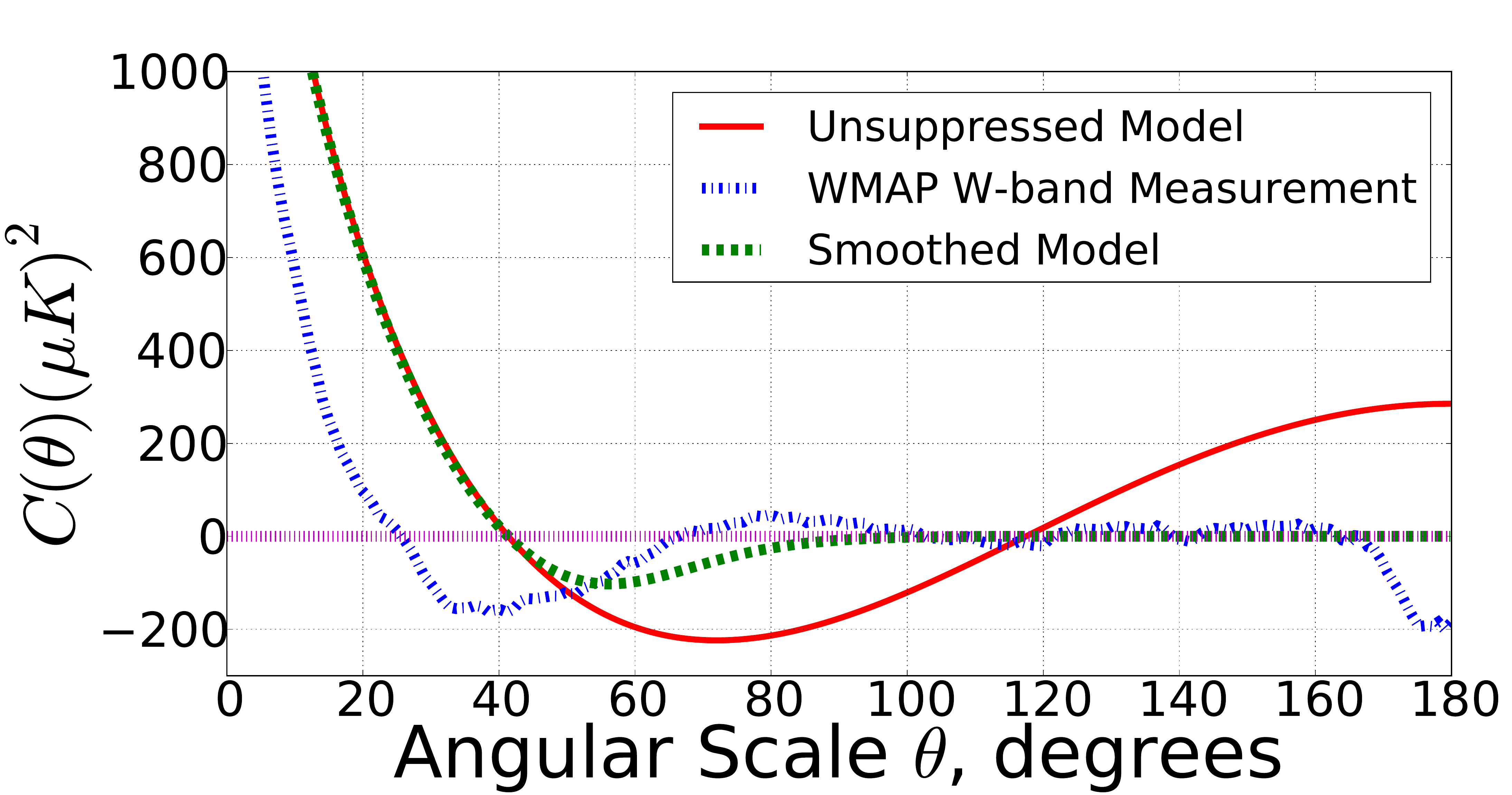}
\caption{The angular two-point function $C(\theta)$. Measurement from
  the W-band WMAP 7-year maps, adopted from \cite{Sarkar}, is shown
  here in blue, while our model in which $C(\theta)$ smoothly
  transitions to zero at higher $\theta$ is shown in green. This
  smoothed model mimics and idealizes the behavior of the measured
  $C(\theta)$ at large scales but follows the \lcdm prediction (red
  curve) on smaller scales, below roughly 50 degrees. The dotted line
  shows zero correlation for reference.}
\label{fig:ctheta}
\end{center}
\end{figure}

Since we are examining the phenomenon of low power on large angles in the CMB,
it is really the $C(\theta)$ data that we wish to be faithful to, so we take
$C(\theta)$ as our starting point rather than $C_{\ell}$. We start from the
pixel-based measurement of the angular correlation function (adopted from
\cite{Sarkar}), which we denote with a tilde, $\tilde C(\theta)$.  In order to
smooth out the noise in the measured $\tilde C(\theta)$, and thereby simplify
the inverse problem somewhat, we use a ``smoothed model'' for $\tilde C(\theta)$ that
is designed to agree with \lcdm at small angular scales while closely matching
the actual WMAP data at larger angular scales. To this end, we take the \lcdm
$C(\theta)$ and modify it so that it smoothly transitions to zero for $\theta$
above roughly 60 degrees (Figure \ref{fig:ctheta}).

Inverting Eq.~(\ref{eq:Ctheta_vs_Cl}), we can determine the angular
power spectrum coefficients $\tilde{C}_{\ell}$ inferred from our (smoothed) pixel-based estimate $\tilde C(\theta)$
\begin{equation}
\tilde C_{\ell} = 2 \pi \int_{-1}^{1} P_{\ell}(\cos \theta) \tilde C(\theta) d(\cos \theta).
\label{eqn:clctheta}
\end{equation}
We are now in a position to directly address the inverse problem in
Eq.~(\ref{eqn:Delta2def}). We solve this numerically by discretizing the
integral
\begin{eqnarray}
\label{eqn:ClDeltakernel_orig}
\frac{\ell (\ell + 1) \tilde C_{\ell}}{2 \pi} &=& \int d(\ln k) 
\left \lbrack T_{\ell}(k) \right \rbrack ^2 \Delta_{R}^2 (k) \\[0.2cm]
& \equiv & \sum_k F_{\ell k} \Delta_R^2(k).
\label{eqn:ClDeltakernel}
\end{eqnarray}
The kernel $F_{\ell k}$ is extracted from CAMB \cite{CAMB}. The basic
strategy here, distilled in diagrammatic form, is to start from
$\tilde  C(\theta)$ to find the corresponding $\Delta_R^2(k)$:
\begin{equation}
\tilde C(\theta) \rightarrow \tilde C_{\ell} \rightarrow \Delta_R^2(k).
\label{eqn:flowchart}
\end{equation}

We attempted two different methods of solving this inverse problem,
explained more fully in Appendix \ref{sec:inverse}. Both methods give
similar results, shown with a sample reconstruction in
Fig.~\ref{fig:inverseproblemresult} in the Appendix. While this result
can only be suggestive (given that we used a smoothed model for
$\tilde C(\theta)$ and given that the inverse problem is
ill-conditioned and underdetermined), it does indicate that a
transition to low/zero power on large angular scales in $\tilde
C(\theta)$ can be explained by suppression at low $k$ in the
primordial power spectrum $\Delta_R^2(k)$. If the transition to zero
power in $\tilde C(\theta)$ occurs at about 60 degrees, as it appears
to do in the WMAP cut-sky data, then this corresponds to power
suppression at scales of $k \lesssim 10^{-3.6} \Mpc^{-1} \approx 3.5
\times 10^{-4} h/\Mpc$.

\section{Suppressed primordial large-scale power}\label{sec:suppressed}

The inverse approach from the previous section and the Appendix shows that the
direct inversion of our smoothed $C(\theta)$ leads to a suppression of $P(k)$
at $\logkc \lesssim -3.5$, but the inversion is very noisy
and non-robust, as expected.  We now change tactics and move to the
alternative approach to Eq.~(\ref{eqn:ClDelta}). Instead of treating this as
an inverse problem, we now parameterize $\Delta_R^2(k)$ and treat this as a
(much more stable) forward problem. We utilize a three-parameter model
parameterizing the suppressed $\Delta_R^2(k)$ with an exponential cutoff:
following \cite{Contaldi} and \cite{Mort_Hu_pklow}, we write

\begin{eqnarray}
\DRsup &=& \left [1 - \beta e^{-(k/k_c)^{\alpha}}\right ] 
A_s (k/k_0)^{n_s - 1} \\[0.2cm] 
&\equiv & S(k; k_c, \alpha, \beta) \DRunsup,
\label{eqn:suppressionparameterized}
\end{eqnarray}
where we have implicitly defined the factor $S(k; k_c, \alpha, \beta)\equiv 1
- \beta \exp(-(k/k_c)^{\alpha}) $ by which the power spectrum is
suppressed. The parameter $k_c$ controls the $k$-value of the transition;
$\alpha$ controls the sharpness of the transition; and the extra parameter
$\beta$, which is not found in \cite{Contaldi} or \cite{Mort_Hu_pklow}, allows
the power spectrum to plateau to a value other than zero at low $k$
($S(k)\rightarrow 1 - \beta$ for $k \rightarrow 0$). Note that this
parameterization has enough freedom to mimic the results of the inversion
shown in Fig.\ \ref{fig:inverseproblemresult} almost perfectly.

\begin{figure}[t]
\begin{center}
\includegraphics[width=.45\textwidth]{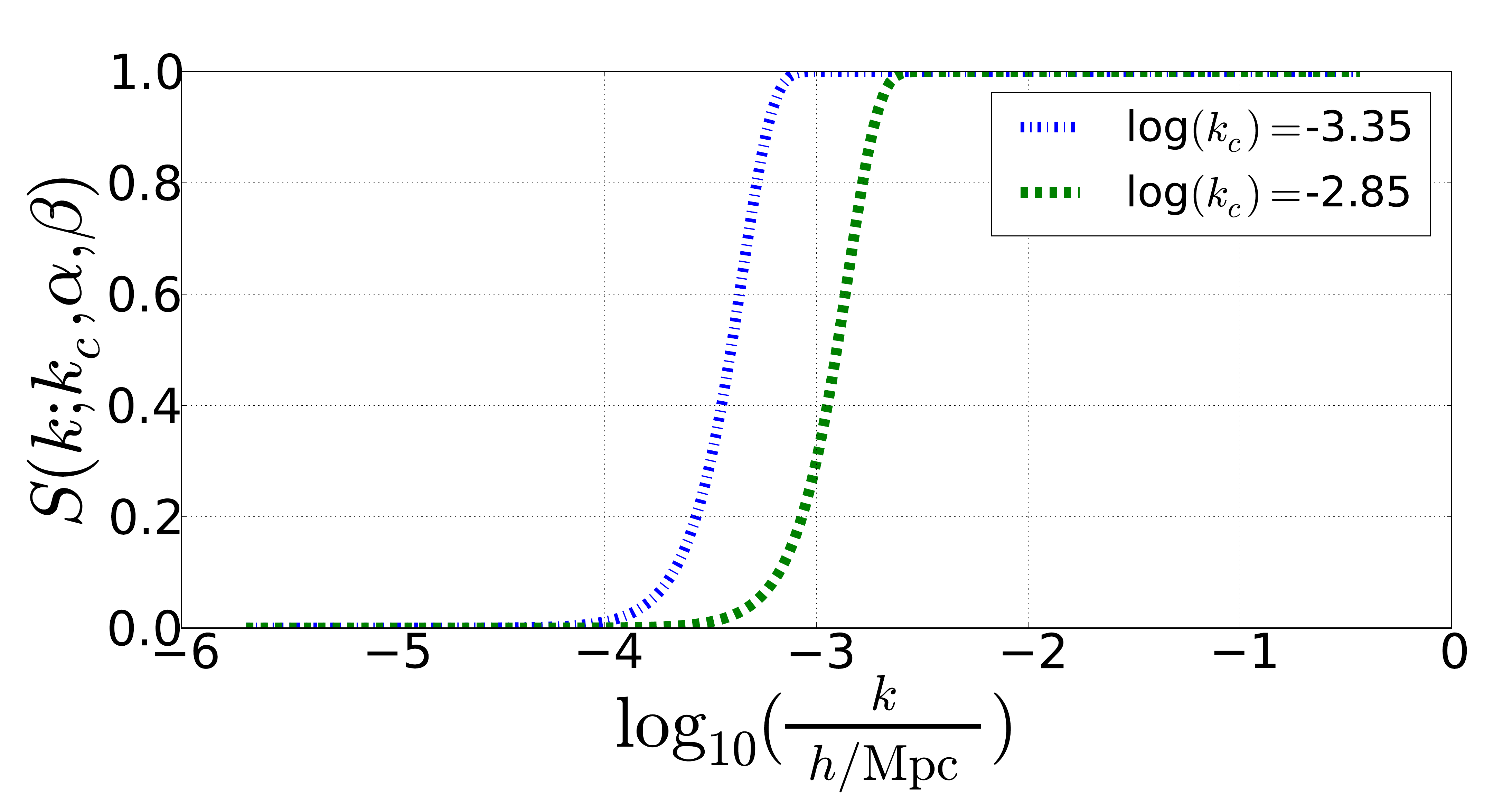}
\includegraphics[width=.45\textwidth]{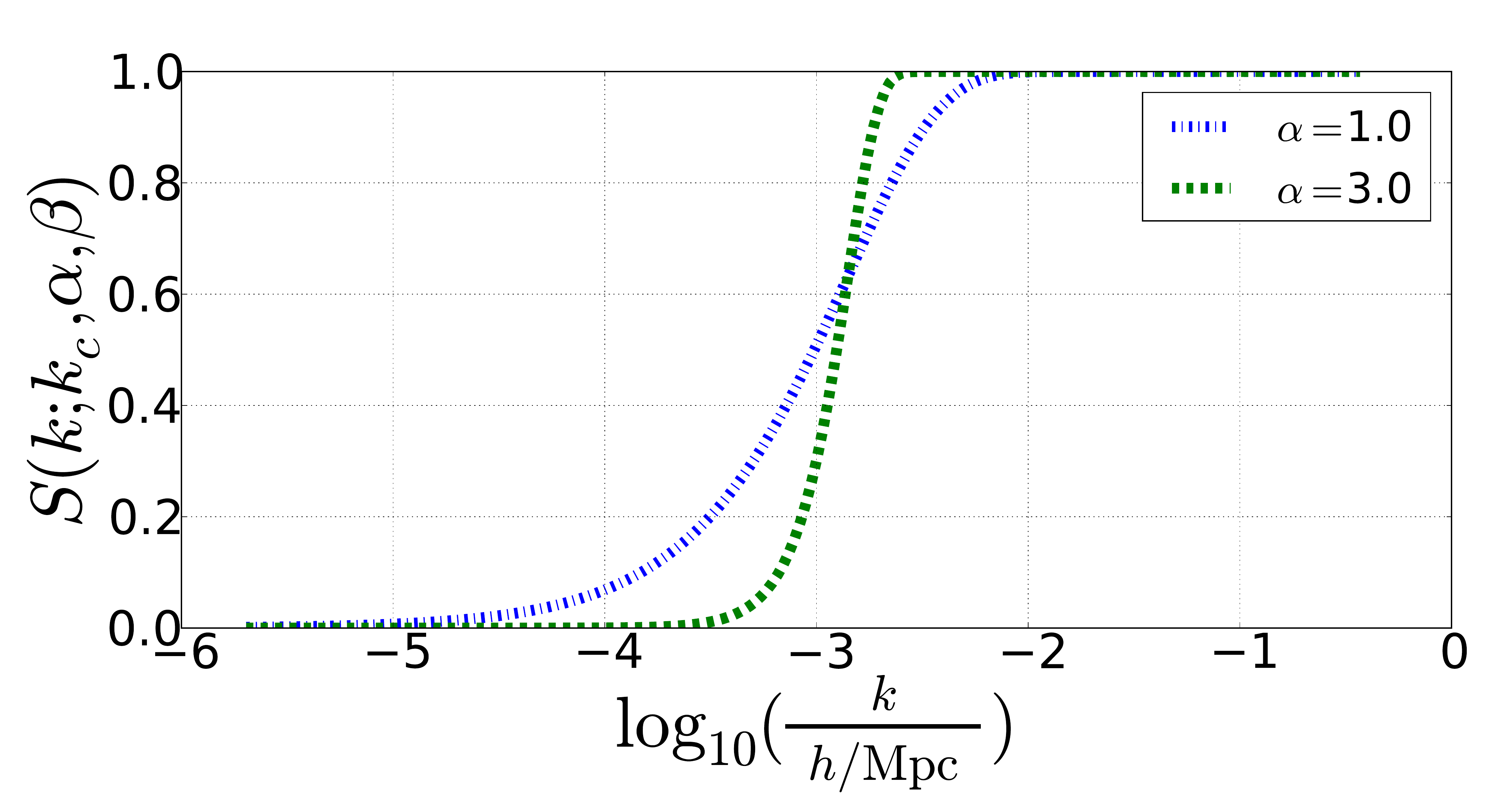}
\includegraphics[width=.45\textwidth]{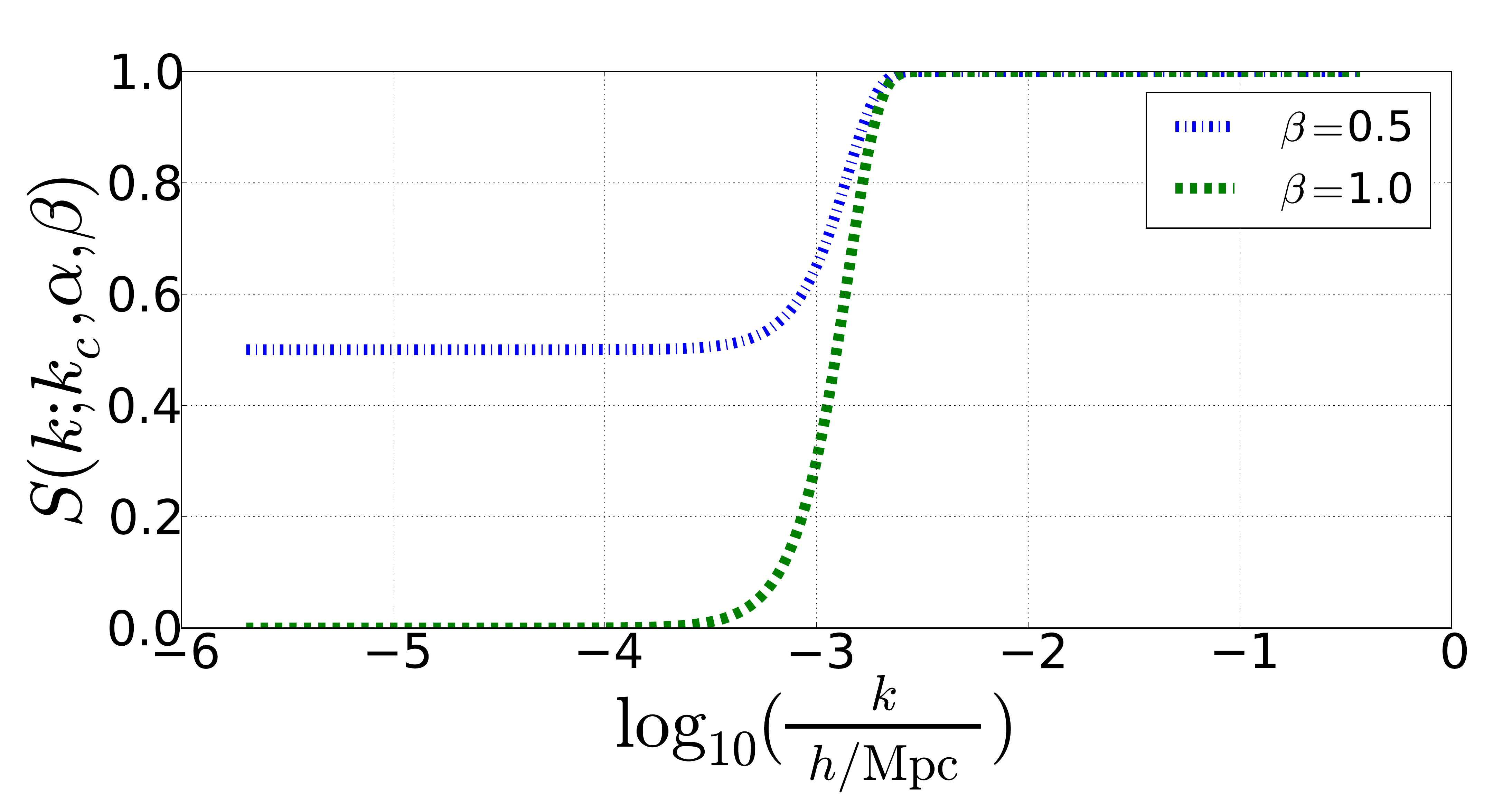}
\caption{Illustration of what the suppression factor $S(k)$ looks like for
  various combinations of parameters. The default parameters are $\log(k_c)
  \equiv \logkc = -2.85$, $\alpha = 3.0$,
  and $\beta = 1.0$. In each plot, one of these parameters is varied, while
  the other two are held fixed at their default values.}
\label{fig:skvsk}
\end{center}
\end{figure}

For a given set of parameters $\{k_c, \alpha, \beta\}$, we have a well-defined
$\Delta_R^2(k)$ and can thus use the numerical kernel $F_{\ell k}$ to find the
corresponding $C_{\ell}$'s, as in Eq.~(\ref{eqn:ClDeltakernel}). We can then
determine which combinations of parameters give $C_{\ell}$'s that fit the
observed WMAP $C_{\ell}$'s, and likewise the observed $C(\theta)$. We are now
moving in a direction opposite the one in Eq.~(\ref{eqn:flowchart}):
\begin{equation}
\Delta_R^2(k) \rightarrow C_{\ell} \rightarrow C(\theta).
\label{eqn:flowchartbackward}
\end{equation}
Plots of the suppression factor $S(k)$ for several sample parameter
values are shown in Figure \ref{fig:skvsk}.

When varying the suppression parameter $k_c$ (and, optionally, $\alpha$ and
$\beta$), we have {\it not} simultaneously varied other parameters that
describe the primordial power spectrum, such as the dark matter and baryon
densities, spectral index, etc. The reason, in addition to simplicity, is that
none of these other parameters can mimic the large-scale suppression of power,
and therefore, power suppression is not degenerate with other cosmological
parameters. The one possible exception would be primordial nongaussianity of
the local type, which does indeed affect the power spectrum of halos (and,
thus, galaxies) on large scales \cite{Dalal}; however, including this
degeneracy is beyond the scope of this project.

\section{Statistical tests}\label{sec:statistical}

\begin{figure}[t]
\begin{center}
\includegraphics[width=.45\textwidth]{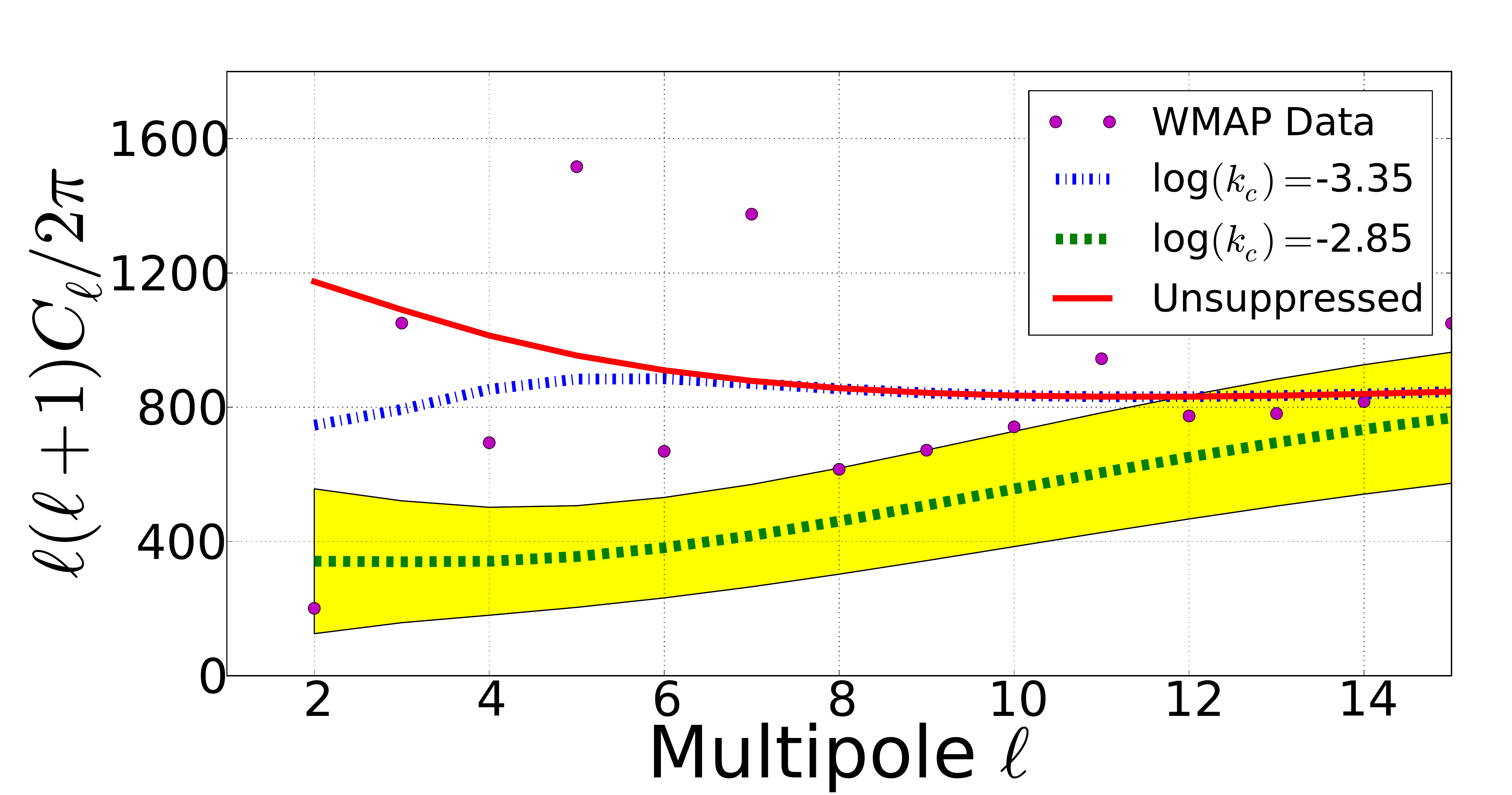}
\includegraphics[width=.45\textwidth]{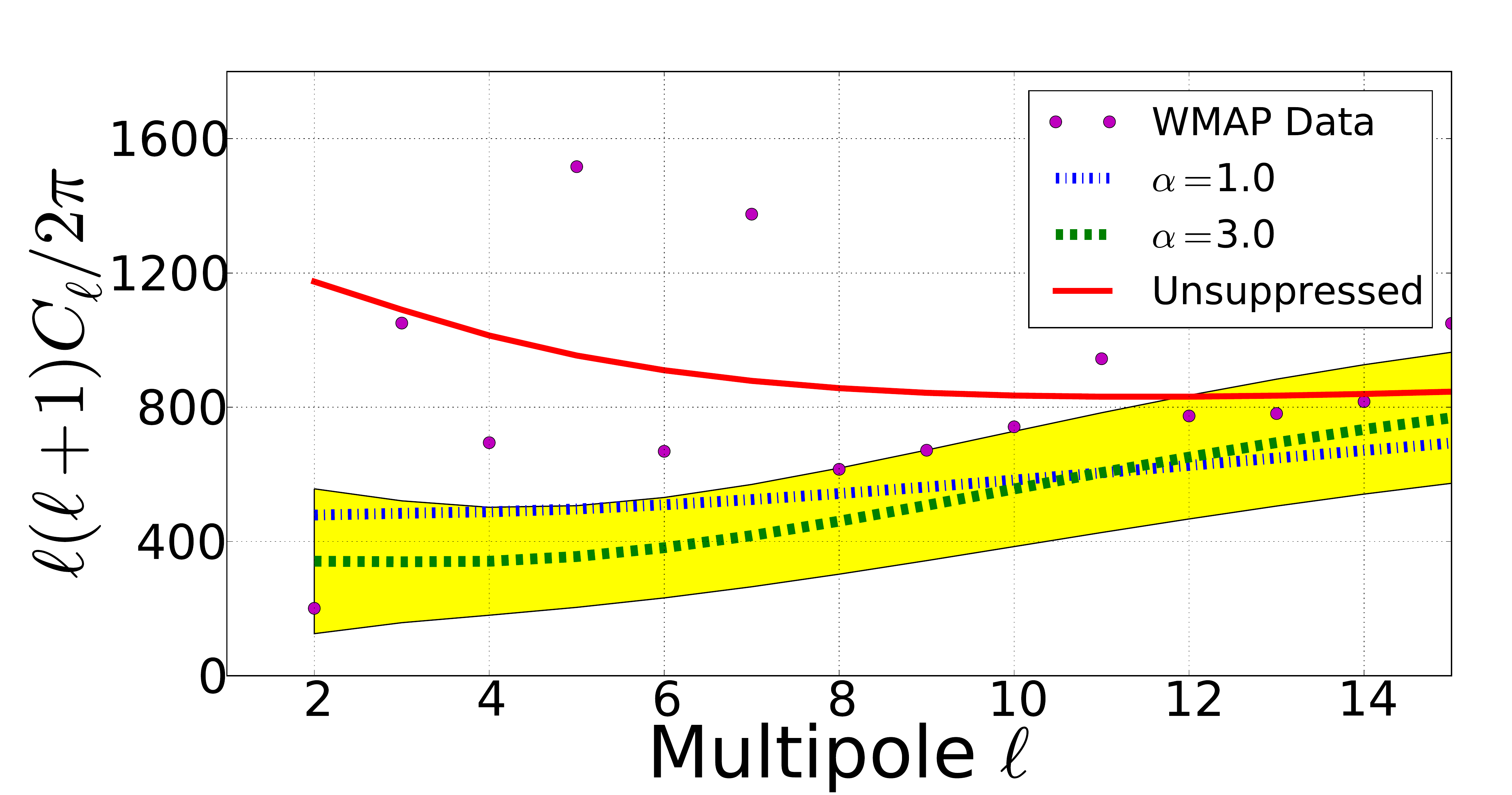}
\includegraphics[width=.45\textwidth]{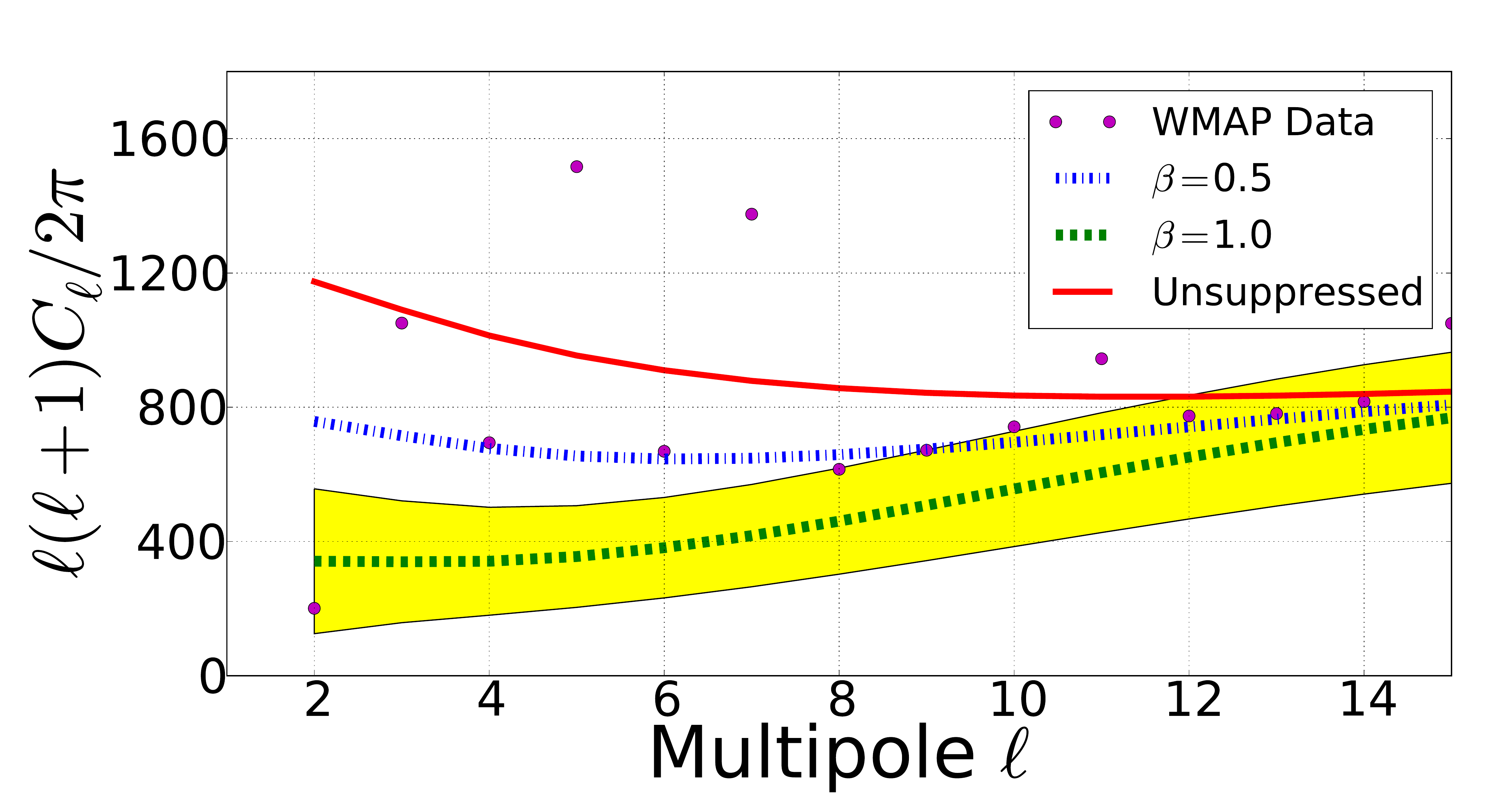}
\caption{Effects of the suppression on the angular power spectrum
  $C_{\ell}$. The default parameters are again $\log(k_c) \equiv
  \logkc = -2.85$, $\alpha = 3.0$, and $\beta = 1.0$. In each panel,
  one of these parameters is varied, while the other two are held
  fixed at their default values. The WMAP measurements are shown as
  points with measurement error bars (too small to see in this
  plot). Cosmic variance (assuming full-sky measurements) is plotted
  as a band around the most heavily suppressed model (green curve) in
  each panel.}
\label{fig:clvary}
\end{center}
\end{figure}

\begin{figure}[t]
\begin{center}
\includegraphics[width=.45\textwidth]{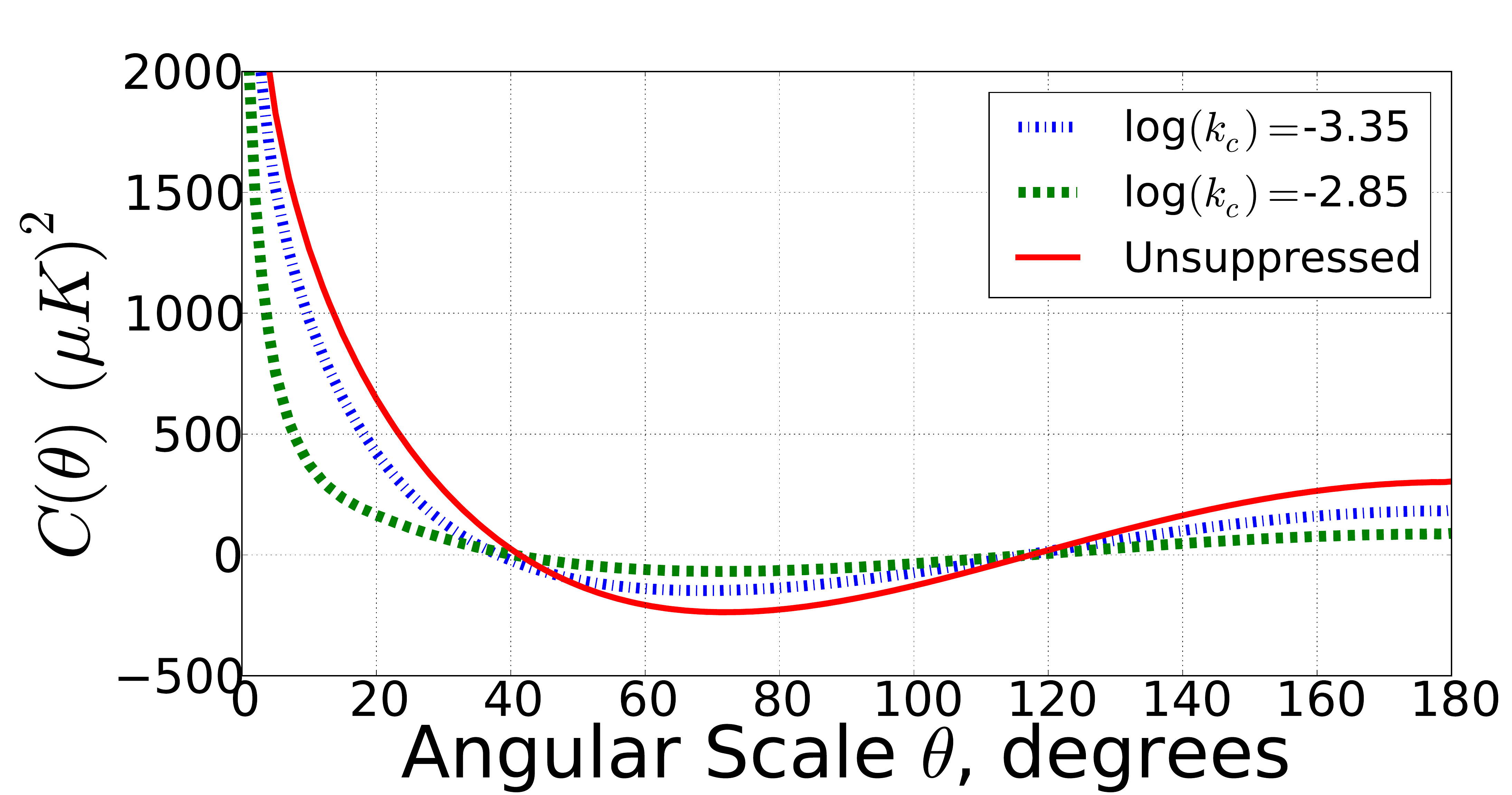}
\includegraphics[width=.45\textwidth]{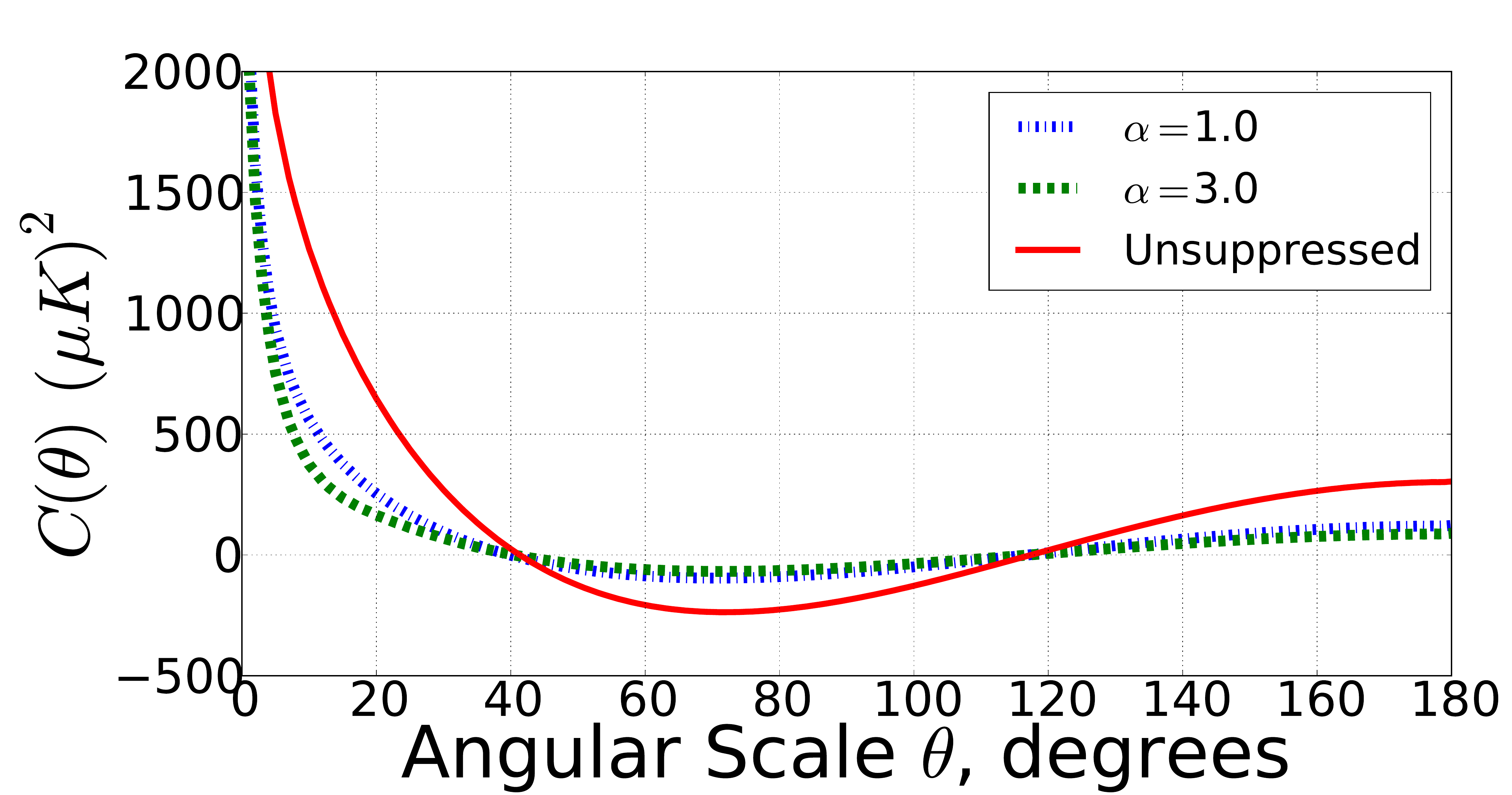}
\includegraphics[width=.45\textwidth]{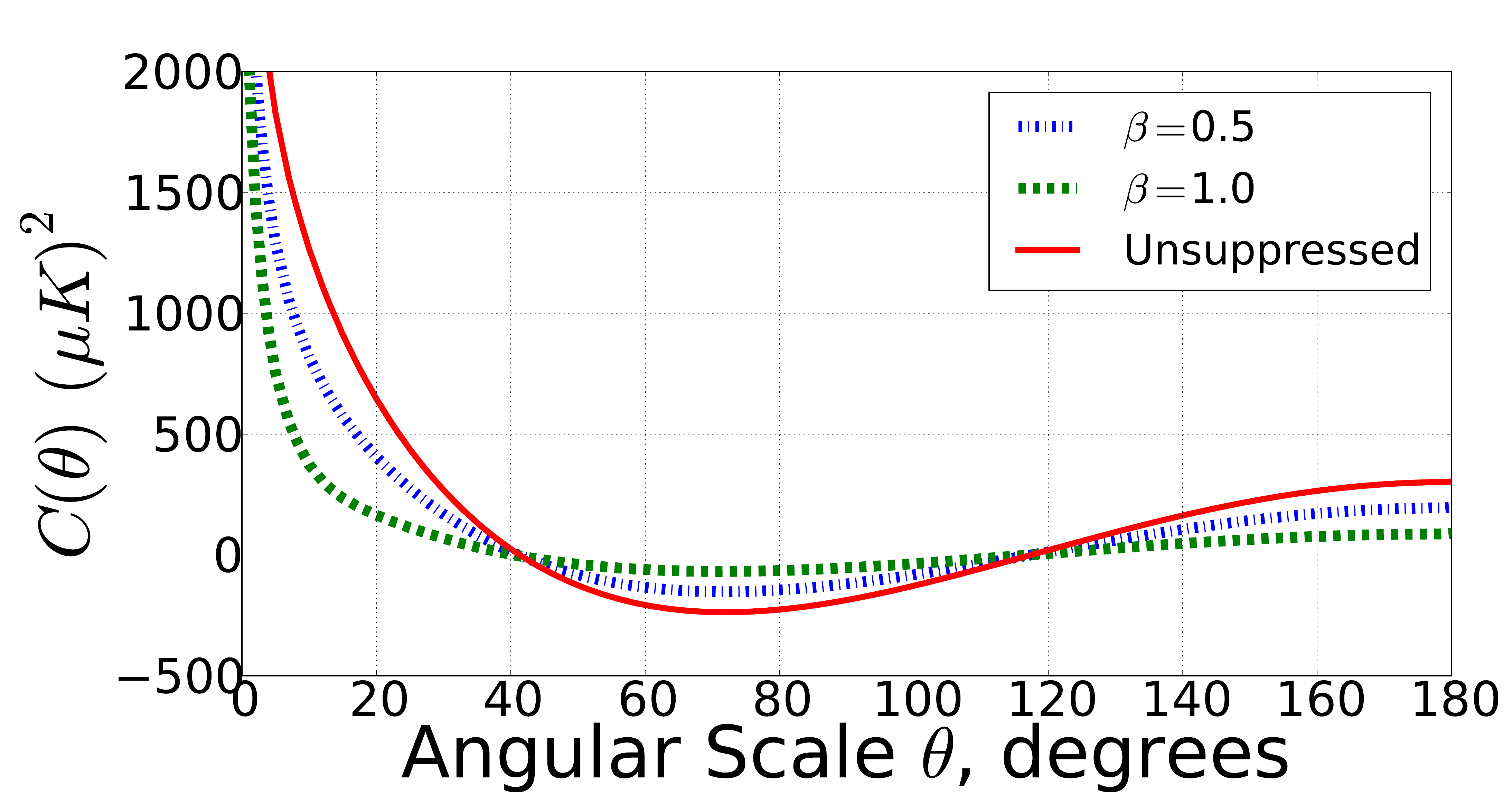}
\caption{ Effects of the suppression on the real-space two-point correlation
    function in the CMB, $C(\theta)$. The default parameters are again
    $\log(k_c) \equiv \logkc = -2.85$, $\alpha = 3.0$, and $\beta = 1.0$. In
    each plot, one of these parameters is varied, while the other two are held
    fixed at their default values.}
\label{fig:cthetavary}
\end{center}
\end{figure}

We are interested in how a suppressed primordial power spectrum, as
given in Eq.~(\ref{eqn:suppressionparameterized}), affects both
$C_{\ell}$ and $C(\theta)$. Figures \ref{fig:clvary} and
\ref{fig:cthetavary} show examples of how these quantities vary with
changes in the parameters.

More specifically, we look to quantify whether, and to what extent, a
suppressed primordial power spectrum gives a better fit to
observations of $C_{\ell}$ (typically inferred using
maximum-likelihood-type techniques at large scales) and $C(\theta)$
estimated on cut-sky maps using a pixel-based estimator. We restrict
attention here to varying $k_c$ rather than all three parameters,
since variations in $k_c$ have the greatest effects on the likelihood,
and also since $k_c$ directly controls the scale at which the
suppression occurs.

In the case of $C_{\ell}$, it is relatively straightforward to quantify the
fit between a given suppressed model and the WMAP data. We simply generate a
suppressed model using CAMB and then feed the resulting $C_{\ell}$ spectrum
into the WMAP likelihood code\footnote{Available from
  \url{http://lambda.gsfc.nasa.gov/product/map/dr4/likelihood_info.cfm}.}  to
obtain a goodness-of-fit criterion. Further details are covered in
Sec. \ref{subsec:cl}.

Quantifying the fit between a suppressed-model $C(\theta)$ and the
WMAP data requires a bit more work. In Sec. \ref{subsec:ctheta}, we
examine the statistic $S_{1/2}$, which gives a measure of correlations
in the CMB above scales of $60\degr$. In analogy to our treatment of
the angular power spectrum, we define a $\chi^2$ statistic to quantify
the goodness of fit between a given suppressed model and the WMAP
data.

\subsection{Angular Power Spectrum $C_{\ell}$}\label{subsec:cl}

To quantify how well the suppressed angular power spectrum $C_{\ell,
  {\rm sup}}$ fits the WMAP observations $C_{\ell, {\rm WMAP}}$, we
use the WMAP likelihood code to compute $\chi^2_{C_{\ell},\rm sup}(k_c) = -2
\ln \mathcal{L}_{\rm sup}(k_c)$ from $C_{\ell, {\rm sup}}$. We
likewise compute $\chi^2_{C_{\ell},\rm unsup} = -2 \ln \mathcal{L}_{\rm unsup}$
from the unsuppressed \lcdm power spectrum $C_{\ell, {\rm unsup}}$,
finding the difference

\begin{equation}
\dccl(k_c) = \chi^2_{C_{\ell},\rm sup}(k_c) - \chi^2_{C_{\ell},\rm unsup}
\label{deltachi2cl}
\end{equation}
as a final quantification of how well the suppressed model fits
$C_{\ell}$ data relative to the unsuppressed \lcdm model. 

A negative $\dccl$ indicates that the suppressed model is a better fit to the
WMAP data than \lcdm. For models with a large amount of suppression (i.e.,
high $k_c$), the suppressed model affects the $C_\ell$ at increasingly smaller
scales (higher $\ell$), making them inconsistent with the $C_\ell$ measured
from WMAP and thus making the quantity $\dccl$ large and positive.

The results are shown further below in Fig.~\ref{fig:pclps12}, with discussion
in Section \ref{sec:constraints}.

\subsection{Angular Correlation Function $C(\theta)$}\label{subsec:ctheta}

We now study to what extent the power suppression from our model can account
for the low correlations observed above $60\degr$ on the cut-sky WMAP maps
\cite{Spergel2003,wmap123,wmap12345}. The statistic $S_{1/2}$, defined in
\cite{Spergel2003}, quantifies the lack of correlation above 60 degrees
\begin{equation}
S_{1/2} \equiv \int_{-1}^{1/2} \left \lbrack C(\theta) \right \rbrack ^2 d(\cos \theta).
\label{eqn:s12definition}
\end{equation}
It is possible to calculate $S_{1/2}$ directly from the $C_{\ell}$'s:
\begin{equation}
S_{1/2} = {1 \over (4 \pi)^2} \sum_{\ell, \ell'} (2 \ell + 1)(2 \ell' + 1) C_{\ell} I_{\ell, \ell'}(1/2) C_{\ell'}.
\label{eqn:s12fromCl}
\end{equation}
For details of how the quantity $I_{\ell, \ell'}$ is calculated, see Appendix
A of the published version of Ref.~\cite{wmap12345}.

We would like to generate statistical realizations of the angular
power spectrum based on the underlying primordial power spectrum
$\Delta_R^2$. To do that, we first calculate the expected angular
power spectrum $C_{\ell, {\rm original}}$ (using
Eq.~(\ref{eqn:ClDeltakernel_orig})), and then create realizations
using
\begin{equation}
C_{\ell, {\rm realization}} = f C_{\ell, {\rm original}}
\label{eqn:realizations}
\end{equation}
where the multiplicative factor
\begin{equation}
f = \frac{\Gamma(k=(2 \ell+1) \fsky/2, \theta=2)}{(2 \ell+1) \fsky}:
\label{eqn:f}
\end{equation}
the numerator is drawn from a gamma distribution with scale parameter
$\theta=2$ and shape parameter $k=(2 \ell+1) \fsky/2$, and the
denominator ensures that the mean of $f$ is unity. The reason we draw
from a gamma distribution is that this is the appropriate
generalization of a $\chi^2$ distribution when the number of degrees
of freedom is non-integer, as it is above for $\fsky \neq 1$;
$\Gamma(k=r/2, \theta=2)$ is identical to a $\chi^2$ distribution for
integer $r$ degrees of freedom. We adopt $\fsky=0.75$ for the
remainder of this paper. [Modeling of the noise is unnecessary since
cosmic variance dominates at these large scales.]

\begin{figure}[t]
\begin{center}
\includegraphics[width=.45\textwidth]{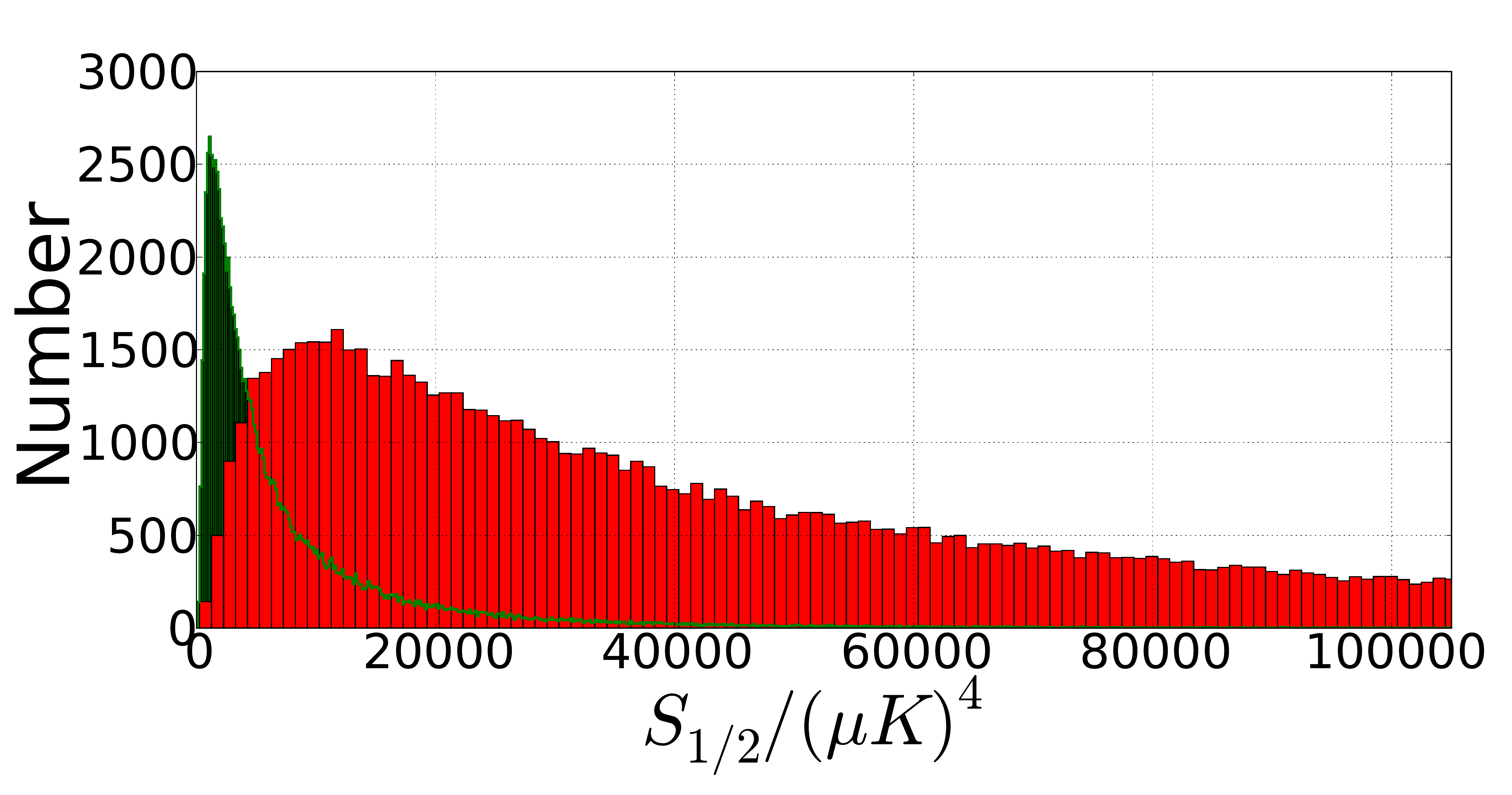}
\includegraphics[width=.45\textwidth]{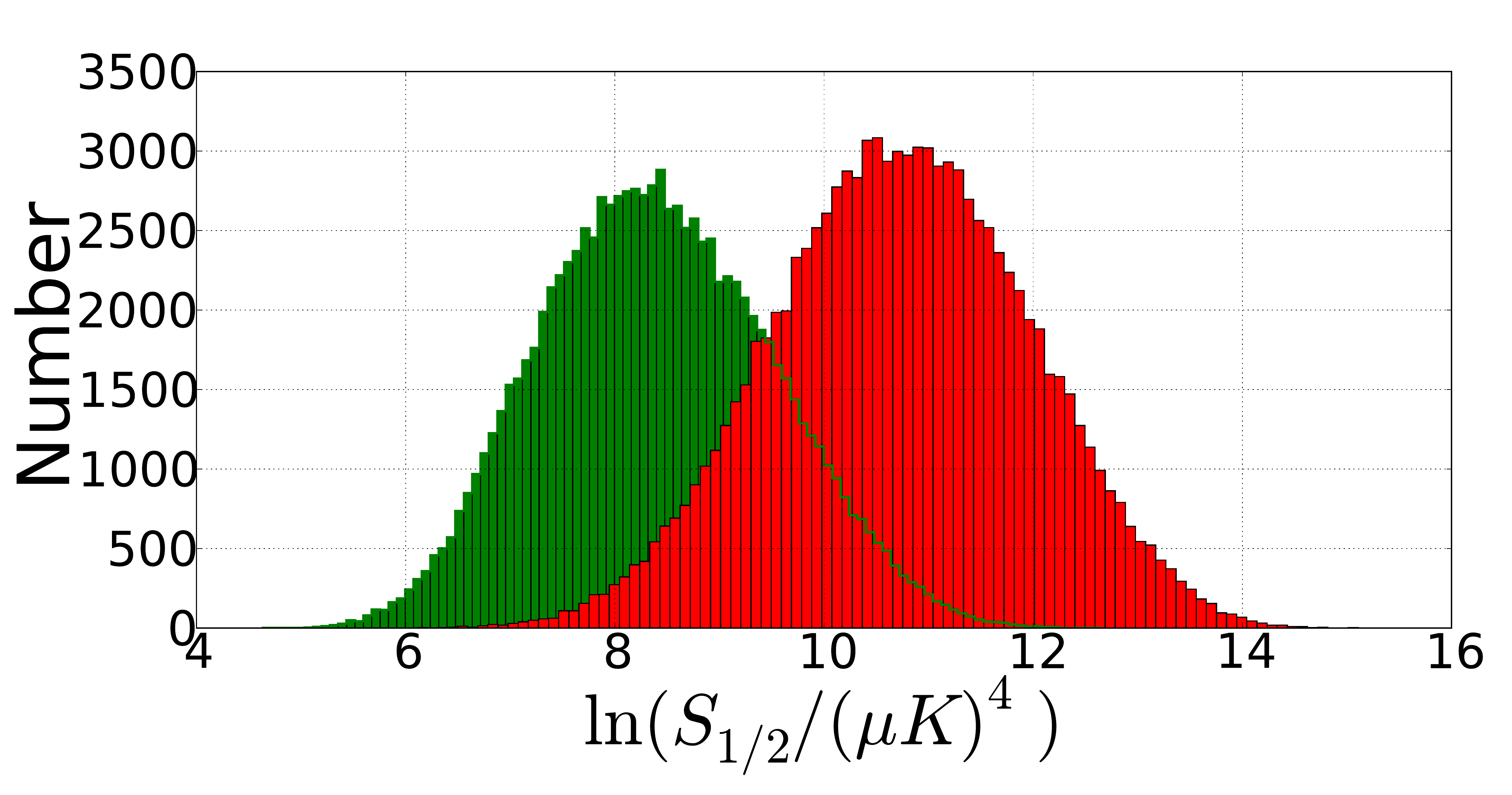}
\caption{The distribution of our realizations of the statistic
  $S_{1/2}$ for the unsuppressed (red histograms) and a sample
  suppressed (green histograms) model. The suppressed model has
  $\logkc = -2.85$, $\alpha = 3.0$, and $\beta=1.0$, and has a better
  fit than the unsuppressed model by $\dcs = -7.9$.  The bottom panel
  clearly shows that the distribution of $S_{1/2}$ is lognormal,
  whether or not the underlying power spectrum is suppressed.}
\label{fig:s12distribution}
\end{center}
\end{figure}

We examine the resulting distribution of $S_{1/2}$ values by performing
100,000 realizations of the $C_{\ell}$ for $2\leq\ell\leq 50$ (going to higher
values of $\ell$ barely changes the $S_{1/2}$ statistic, since scales above
$60\degr$ are mostly affected by $\ell\lesssim 10$), assuming central values
$C_{\ell, {\rm original}}$ calculated based on the suppressed primordial power
spectrum as in Eq.~(\ref{eqn:ClDeltakernel}), and calculating $S_{1/2}$ for
each set of $C_{\ell, {\rm realization}}$. We find that $S_{1/2}$ is
distributed approximately according to a lognormal distribution, both for
suppressed and unsuppressed models, and regardless of the particular value of
$k_c$. This is illustrated in Fig.~\ref{fig:s12distribution}.

In the sample suppressed model ($\logkc = -2.85$) shown in
Fig.~\ref{fig:s12distribution}, the histogram of $S_{1/2}$ peaks at
$1000 (\mu K)^4$, has a mean of $8300 (\mu K)^4$, and a median of
$4300 (\mu K)^4$. The histogram of $\ln(S_{1/2})$ peaks at 8.4,
corresponding to an $S_{1/2}$ of $4400 (\mu K)^4$. These values are
all much lower than the mean value expected in the best-fit \lcdm
cosmology (about $50,000 (\mu K)^4$), but bigger than the value
measured in WMAP cut-sky maps (about $1000 (\mu K)^4$).

In order to calculate a $\chi^2$ statistic in analogy to the $\dccl$ above, we
first transform the lognormal $S_{1/2}$ distribution to a Gaussian by taking
the natural log of the $S_{1/2}$ values. The result, a nearly perfect
Gaussian, is shown for the unsuppressed and the sample suppressed model in the
lower panel of Fig.~\ref{fig:s12distribution}. We can then calculate the
$\chi^2$ corresponding to the probability of getting a certain value
$S_{1/2}^{\rm obs}$ of $S_{1/2}$
\begin{equation}
\chi^2_{S_{1/2}, {\rm sup}} = \left \lbrack \frac{\ln (S_{1/2}(k_c)) - \ln (S_{1/2}^{\rm obs})}
{\sigma_{\ln(S_{1/2})}} \right \rbrack ^2
\label{eqn:chi2s12}
\end{equation}
where $\ln(S_{1/2}(k_c))$ is the mean over the realizations of $\ln(S_{1/2})$
for the given $k_c$ and $\sigma_{\ln(S_{1/2})}$ is the standard deviation over
all realizations. For the purposes of this paper, we choose $S_{1/2}^{\rm obs}
= 1000\, (\mu$K$)^4$, since this is (roughly) the value of $S_{1/2}$ favored
by the cut-sky WMAP observations \cite{Spergel2003,wmap123,wmap12345,Sarkar}.

We also performed 6,500,000 realizations of the $C_{\ell}$ assuming
central values $C_{\ell, {\rm original}}$ corresponding to the
\textit{unsuppressed} \lcdm model. From the $S_{1/2}$ values
associated with these realizations, we calculate $\chi^2_{S_{1/2},
  {\rm unsup}}$ in exact analogy to Eq.~(\ref{eqn:chi2s12}), and then
compute
\begin{equation}
\dcs(k_c) = \chi^2_{S_{1/2},\rm sup}(k_c) - \chi^2_{S_{1/2},\rm unsup}.
\label{deltachi2s12}
\end{equation}

A combined statistic that takes into account both the measurements of the
angular power spectrum $\Cl$ and the total angular correlation above $60\degr$
parameterized by the statistic $S_{1/2}$, is then given by\footnote{Since we
  are interested in how likely the low value of $S_{1/2} \approx 1000 (\mu
  K)^4$ is, given suppression of power, we could have simply calculated
  $P(S_{1/2} < 1000)$ -- the probability that $S_{1/2}$ is \textit{as low as}
  1000 -- instead of performing the more complicated calculation above to
  obtain $P(S_{1/2})$. However, the danger in doing this is that suppression
  on small enough scales leads to values of $S_{1/2}$ that are much lower than
  1000, and then the probability that $S_{1/2}$ is \textit{as high as} 1000
  should become low.
  Considering the Gaussian likelihood in $\ln (S_{1/2})$, as we have done,
  correctly penalizes values of $S_{1/2}$ that are too low {\it or} too high.
}

\begin{eqnarray}
  \mathcal{L}(k_c) &=&   
\exp\left (-\frac{\dccl(k_c)}{2}\right )\times 
\exp\left (-\frac{\dcs(k_c)}{2}\right )\nonumber \\[0.2cm]
&\equiv & \pcl \times \pss.
\label{eqn:likelihoods}
\end{eqnarray}
Both $\pss$ and $\pcl$ are normalized so that their values for the
unsuppressed \lcdm model are 1. Hence $\pss$ and $\pcl$ should be interpreted
as the improvement (relative to fiducial unsuppressed $\Lambda$CDM) in how
well a given suppressed model fits the WMAP data for $C_{\ell}$ and
$S_{1/2}$. 
Note that we are not taking the correlation between the
(maximum-likelihood) $C_\ell$ and the (pixel-based) $S_{1/2}$ into
account in our statistic $\mathcal{L}(k_c)$. We define the statistic
in the simplest possible way, by multiplying the individual
likelihoods in $C_\ell$ and $S_{1/2}$. This simple combination is
sufficient, since it favors suppression on scales between the scales
which $\pss$ and $\pcl$ independently prefer (this is confirmed in
Fig.~\ref{fig:pclps12}), and thus captures the essence of how these
two quantities jointly favor suppression.
Note that the main results of this paper, presented in Sec.~\ref{sec:future},
do not depend on exactly how we combine the likelihood in the measured
full-sky $C_{\ell}$ and cut-sky $S_{1/2}$; we only use $\mathcal{L}(k_c)$ from
Eq.~(\ref{eqn:likelihoods}) to get a rough idea about the scale at which
suppression is favored by the data if both measurements are taken at face
value. We then proceed in Sec.~\ref{sec:future} to study the detectability of
suppressed power corresponding to a range of values of the suppression scale
$k_c$; these results do not depend on how $\pcl$ and $\pss$ are combined.

We are now in a position to determine which values of the suppression scale
$k_c$ improve the joint fit to the angular power spectrum $C_\ell$ and the
cut-sky measurements of the angular correlation function $C(\theta)$ above
$60\degr$ (quantified by the statistic $S_{1/2}$).

\section{Current constraints from the CMB}\label{sec:constraints}

\begin{figure}[t]
\begin{center}
\includegraphics[width=.45\textwidth]{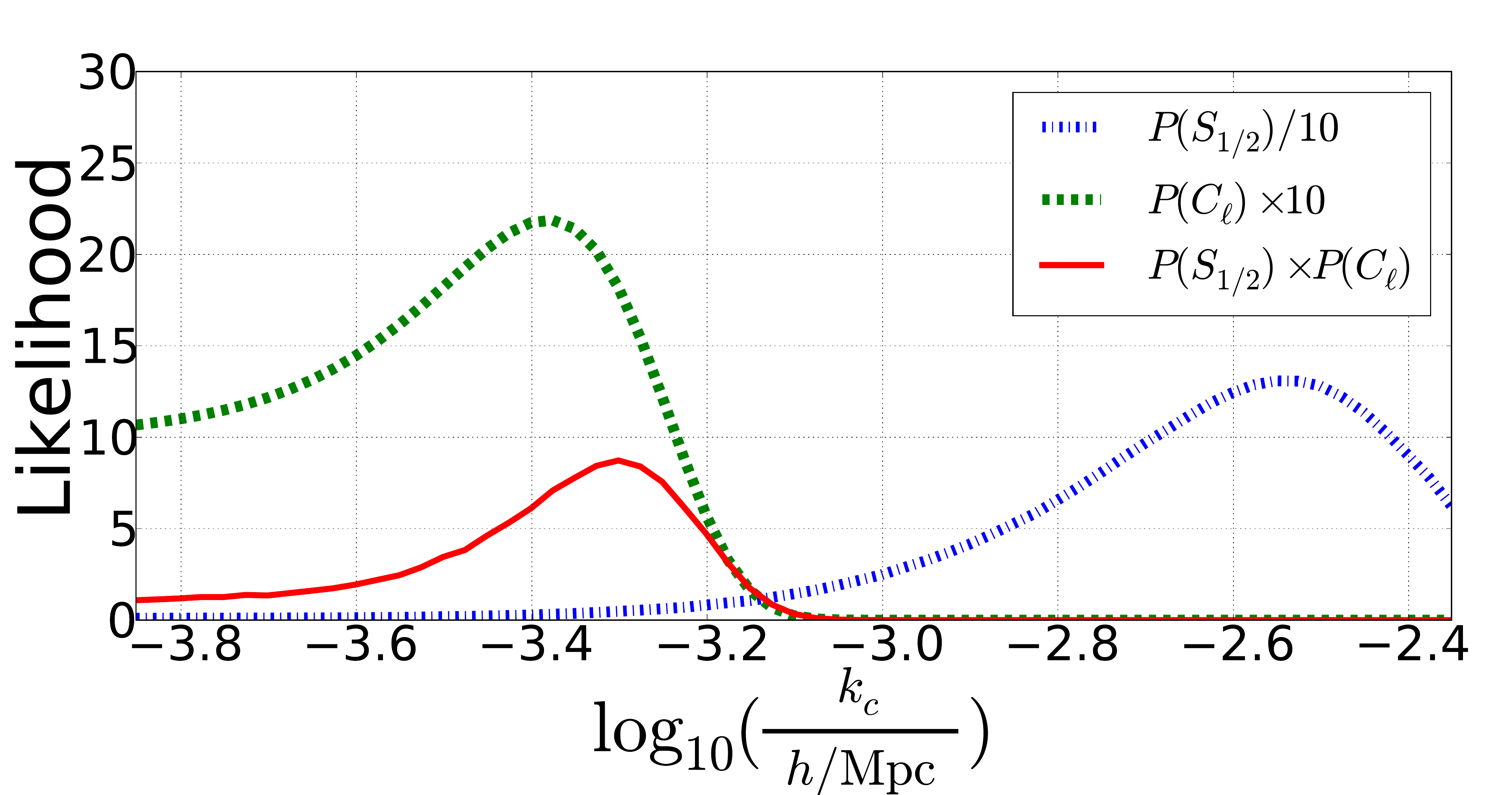}
\includegraphics[width=.45\textwidth]{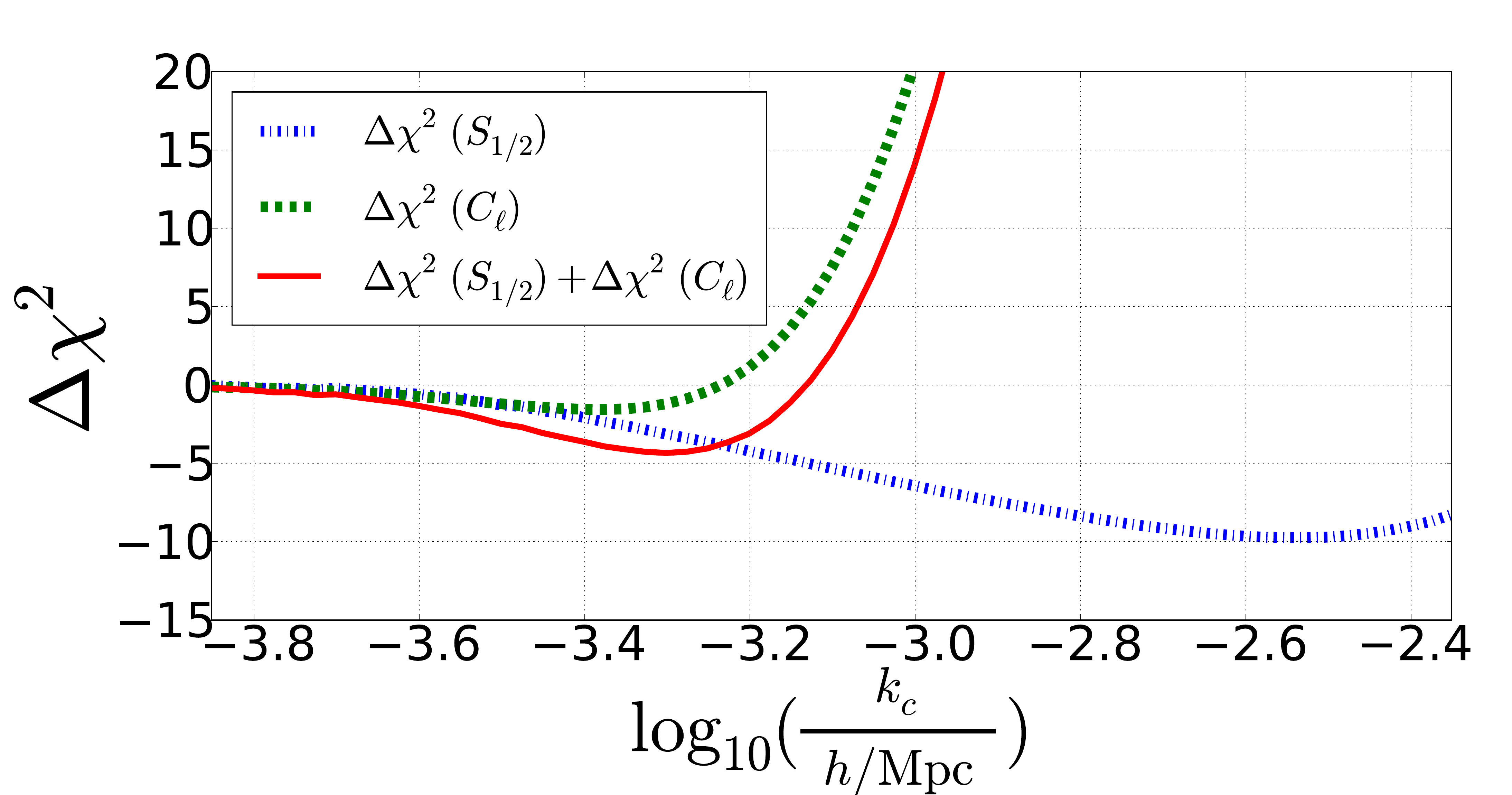}
\caption{The improvements in $\pss$, $\pcl$, and the product thereof
  -- all relative to the unsuppressed model -- are plotted as a
  function of $k_c$. The top panel shows $\pss$ divided by 10 and
  $\pcl$ multiplied by 10. The $\pss$ values are based on 100,000
  realizations of the $C_{\ell}$'s as described in the text. The
  bottom panel shows the same information as the top panel, but puts
  it in terms of chi-square goodness-of-fit statistics. The maximum
  improvement in $\pss \times \pcl$, relative to the unsuppressed
  \lcdm model, is a factor of 8.7 ($\dcs + \dccl = -4.3$), occurring
  at $\logkc \approx -3.3$. All calculations were performed assuming
  $\fsky = 0.75$.}
\label{fig:pclps12}
\end{center}
\end{figure}


The top panel of Fig.~\ref{fig:pclps12} shows $\pss$, $\pcl$, and their
product as a function of $k_c$, with $\alpha$ held fixed at 3.0 and $\beta$
held fixed at 1.0. The bottom panel displays the same result using
$\chi^2_{S_{1/2}}$ and $\chi^2_{C_{\ell}}$ on the vertical axis instead of
$\pss$ and $\pcl$.

As indicated in Fig.~\ref{fig:pclps12}, introducing suppression in the
primordial power spectrum can increase the likelihood of both the observed
$C_{\ell}$ and the observed $S_{1/2}$, but these two observations favor
suppression at different scales. The likelihood of the $C_{\ell}$ is improved
by at most a factor of
2.2, with the improvement peaking at 
$\logkc = -3.4$, while greater suppressions can improve the likelihood of the
$S_{1/2}$ data by huge factors of up to 131, peaking around 
$\logkc = -2.6$ (note the plotting scale of likelihoods in
Fig.~\ref{fig:pclps12}, where individual likelihood curves are divided
or multiplied by 10 for visual clarity). The $C_{\ell}$ measurements
thus favor suppression on very large scales, while the cut-sky
$S_{1/2}$ favor suppression all the way down to relatively small
scales, where suppression is overwhelmingly ruled out by $C_{\ell}$
data. This is another reminder of just how low the pixel-based cut-sky
measurement of $S_{1/2}$ is. It is also a reminder of the fact that
such a low value of $S_{1/2}$ represents a conspiracy of the
low-$\ell$ $C_{\ell}$ values: the WMAP cut-sky data indicate that
$S_{1/2}$ is sufficiently low as to strongly (by factors of over 100
in likelihood) favor suppression of primordial power at scales
corresponding to $k = 10^{-2.6} \approx 0.003\, h/\Mpc$, even though
the maximum-likelihood $C_{\ell}$ favor suppression only weakly, at
far larger scales, and overwhelmingly reject the possibility of
suppression at the scales favored by the cut-sky $S_{1/2}$.  A sky
with such a low $S_{1/2}$ as the WMAP cut sky \textit{ought}
to have $C_{\ell}$'s that are even more suppressed
than the most suppressed model in Fig.~\ref{fig:clvary}. What we see instead
are low-$\ell$ $C_{\ell}$ values that are not so close to zero, but which
instead conspire with one another in just such a manner as to produce an
exceptionally low value of $S_{1/2}$ anyway \cite{wmap12345}.

In any case, we have calculated the product statistic $\mathcal{L}(k_c) = \pss
\times \pcl$, or alternatively $\dcs + \dccl$, as a measure of how
well a given suppressed model fits the WMAP data in both $C_{\ell}$
and $C(\theta)$ (the latter via the specific statistic
$S_{1/2}$). Since $C_{\ell}$ and $S_{1/2}$ data favor suppression at
such different scales, there should be a ``sweet spot'' somewhere
between the peak in $\pcl$ and the peak in $\pss$, where suppression
is moderately favored by both $C_{\ell}$ and $S_{1/2}$, or heavily
favored by one and still allowed by the other. This is indeed what we
find, as indicated by the red curve in Fig.~\ref{fig:pclps12}. Because
suppression on overly small scales (below $\logkc \sim -3.2$) brings
the $C_{\ell}$ data for the suppressed model into severe conflict with
the WMAP $C_{\ell}$'s, the peak of the $\mathcal{L}$ curve occurs
above these scales, even in spite of the huge gains in likelihood that
$\pss$ gives us at much smaller scales, where the gain in $\pss$ is
still substantial and the $C_{\ell}$ data still favor -- or at least
do not heavily disfavor -- suppression. The maximum improvement
possible in $\pss \times \pcl$, relative to the unsuppressed \lcdm
model, is a factor of 8.7 ($\dcs + \dccl = -4.3$), occurring at
$\logkc \approx -3.3$.

The WMAP likelihood code uses a Bayesian (Gibbs sampler) maximum
likelihood method (e.g.\ \cite{Efstathiou2003}) to compute the
fiducial $C_{\ell}$'s at the multipoles $\ell \leq 32$
\cite{Dunkley_wmap5,Larson_wmap7}. We experimented with running the
likelihood code using pseudo-$C_{\ell}$ estimates at low
multipoles\footnote{We did this by turning off the {\tt use\_lowl\_TT}
  option in the {\tt test.F90} routine of the WMAP likelihood code,
  and also switching off polarization by turning off the {\tt use\_TE}
  and {\tt use\_lowl\_pol} options.} and discovered that in this case,
suppression is much \textit{more} heavily favored by the $C_{\ell}$
likelihood than it is in the (presumably more accurate) Gibbs sampler,
or else a similar Maximum Likelihood Estimate (MLE) method. This
result is expected, and holds because the $C_{\ell}$'s that result
from the pseudo-$C_{\ell}$ estimates are lower than those found using
the Gibbs sampler method (see e.g.\ Fig.~15 in \cite{Hinshaw:2006ia}),
and suppression fits them better.  In this case we can get $\dccl$ as
low as $-7.6$, corresponding to improvements in $\pcl$ by factors of
up to 44 (as opposed to roughly 2 in the best-case scenarios discussed
above).

\section{Future detectability using galaxy surveys}\label{sec:future}

The results of the previous section indicate that suppression of primordial
power on large scales can increase the likelihood of both the observed
$C_{\ell}$ angular power spectrum and the observed cut-sky value of $S_{1/2}$,
provided the suppression ``kicks in'' on appropriate scales. Now we turn to
the question of whether large-scale suppressed power could be detected in the
matter power spectrum as measured by upcoming redshift surveys such as the
Large Synoptic Survey Telescope (LSST; \cite{LSST}). If the zero-correlation
signature of large-angle $C(\theta)$ in the CMB is an authentic effect
indicating a deficit of power on the Universe's largest scales, is it possible
to cross-check and verify this result using large-scale-structure data?

Given suppression of the primordial power spectrum $\Delta_R^2(k)$ as
parameterized in Eq.~(\ref{eqn:suppressionparameterized}), the matter power
spectrum will be suppressed by the same factor as $\Delta_R^2(k)$:
\begin{equation}
P_{\rm sup}(k) = S(k) P_{\rm unsup}(k)
\label{eqn:pksuppressed}
\end{equation}
where $S(k) \equiv S(k; k_c, \alpha, \beta)$ is the same as before.  We wish
to determine whether this suppressed matter power spectrum could be
distinguished from the unsuppressed \lcdm matter power spectrum $P_{\rm
  unsup}(k)$ by a large-volume redshift survey.

When measuring the matter power spectrum with a redshift survey, the error
bars in each thin slice in redshift $dz$ and wavenumber $dk$ are given
by the Feldman-Kaiser-Peacock (FKP; \cite{FKP}) formula
\begin{equation}
\sigma_P^2(k,z) = \frac{4 \pi^2 P(k,z)^2}{k^2\, d k\, d V_{\rm eff}}
\label{eqn:errorbars}
\end{equation}
where the effective volume element $dV_{\rm eff}(k,z)$ is related to a 
comoving volume element via
\begin{equation}
  d V_{\rm eff}(k,z) = \left \lbrack \frac{n(z)P(k,z)}{1+n(z)P(k,z)} 
  \right \rbrack ^2 dV_{\rm survey}(z).
\label{eqn:Veff}
\end{equation}
The differential survey volume is given in terms of $dz$ via
\begin{equation}
dV_{\rm survey} = \Omega_{\rm   survey} \frac{r(z)^2}{H(z)} dz,
\end{equation} 
where $r(z)$ is the comoving distance as a function of redshift, $H(z)$ is the
Hubble parameter as a function of redshift, and $\Omega_{\rm survey}$ is the
angular size of the survey in steradians. 

The number density of galaxies $n(z)$ can be found from
\begin{equation}
  n(z) = m(z) \times \frac{N_{\rm tot}}{\Omega_{\rm survey} \int m(z) [r(z)^2/H(z)] dz}
\label{eqn:nz}
\end{equation}
where the second term on the right-hand side provides a normalization. Here
$N_{\rm tot}$ is the total number of galaxies in the survey and $m(z)$ is the
(unnormalized) number density of galaxies, whose functional form we adopt to
be
\begin{equation}
m(z) = \frac{z^2 e^{-z/z_0}}{2 z_0^3}.
\label{eqn:mz}
\end{equation}
We take $z_0 = 0.35$, corresponding to the density roughly expected in the
imaging portion of the LSST survey \cite{HTBJ}, and assume a 23,000 square
degree redshift survey with $0.5$, $5$ or $50$ spectra per square
arcminute. [Note that the $0.5$ and $5$ gal/arcmin$^2$ cases are realistic,
  being targeted by surveys in the near future \cite{BigBoss,Wang_space},
  while $50$ gal/arcmin$^2$ corresponds to the more aggressive case where
  spectra of most galaxies in the imaging portion of the survey are taken.]

Given a suppressed power spectrum as in Eq.~(\ref{eqn:pksuppressed}),
we can calculate
\begin{eqnarray}
  d \chi^2 &\equiv & \frac{\left \lbrack P_{\rm unsup}(k,z) - P_{\rm sup}(k,z) 
\right \rbrack ^2}{\sigma_P^2} \\[0.2cm] 
& = & \left \lbrack \frac{nP}{1+nP} \right \rbrack ^2 
  \left \lbrack \frac{\Omega_{\rm survey} k^3}{4 \pi^2 P^2} \right \rbrack 
  \left \lbrack \frac{r^2}{H} \right \rbrack \nonumber \\[0.2cm] 
&&\times \left \lbrack P_{\rm unsup} - P_{\rm sup} \right \rbrack ^2 d(\ln k) dz
\label{eqn:dchi2}
\end{eqnarray}
and then integrate in order to find the $\chi^2$ statistic for how
well the survey can distinguish between the suppressed and
unsuppressed models. 

Note that in the above two equations, wherever a $P \equiv P(k,z)$ occurs
without being marked as either suppressed or unsuppressed, this is intended to
indicate that either $P_{\rm sup}$ or $P_{\rm unsup}$ may be used. Whether we
use $P_{\rm sup}$ or $P_{\rm unsup}$ depends entirely on which question we are
trying to answer: If we use suppressed-model error bars, then this $\chi^2
\equiv \chi^2_{\rm sup}$ indicates at what confidence level the survey can
rule out suppression. Meanwhile, if we use unsuppressed-model error bars, then
$\chi^2 \equiv \chi^2_{\rm unsup}$ indicates at what confidence level the
survey can rule out the unsuppressed \lcdm model. Ruling out \lcdm is
considerably more ambitious than ruling out suppression, since the error bars
tend to be smaller when they are based on the suppressed model (due to the
fact that $\sigma_P \propto P+1/n$).

\begin{figure}[b]
\begin{center}
\includegraphics[width=.50\textwidth]{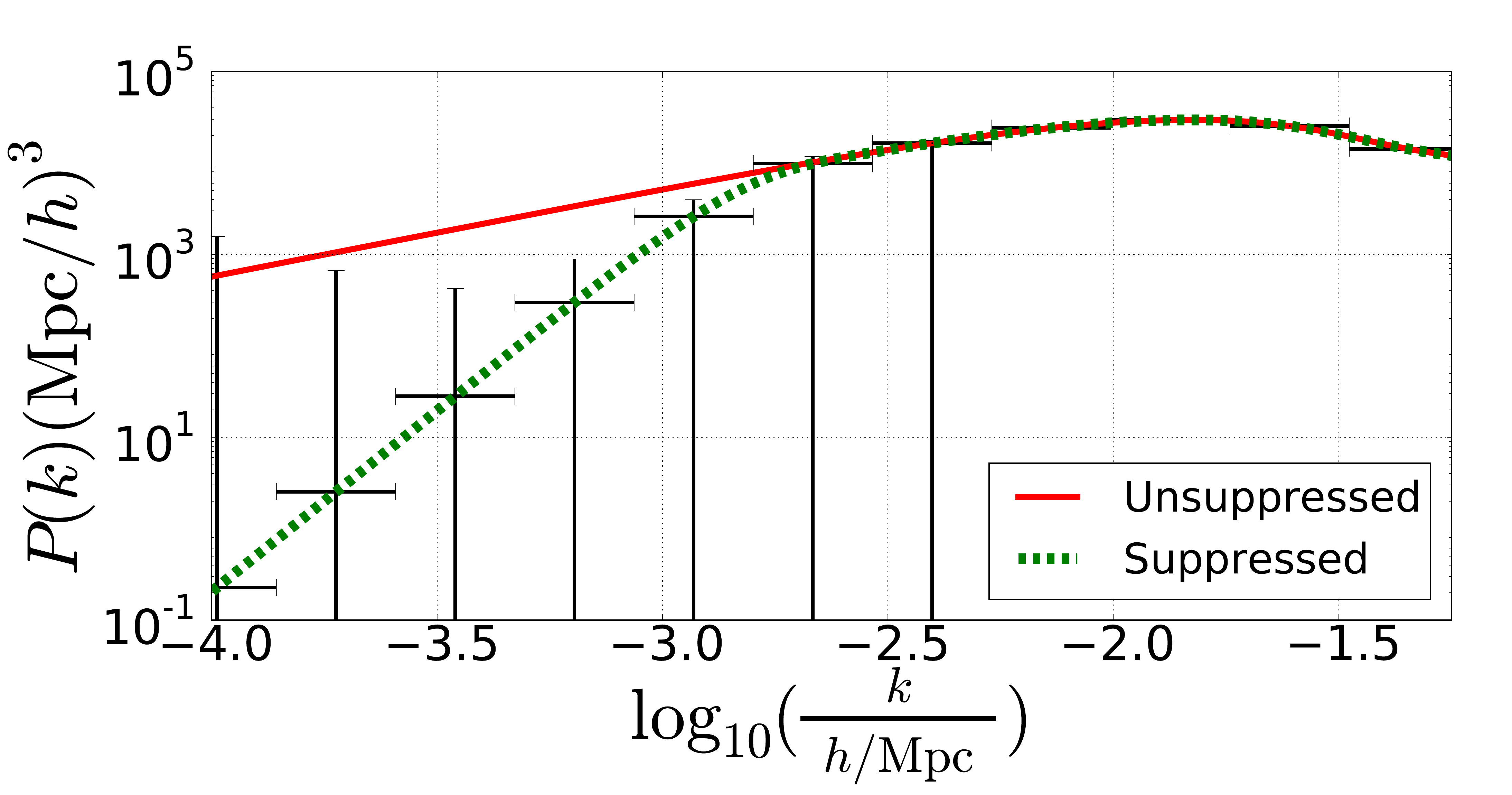}
\caption{Matter power spectrum $P(k)$ with and without suppression. In
  the suppressed model, the parameters are $\logkc = -2.85$, $\alpha =
  3.0$, and $\beta = 1.0$. The power spectrum is shown at $z=0$ for a
  redshift survey with 50 galaxies per square arcminute. The error
  bars are based on the suppressed power spectrum, which means that a
  sufficiently high $\chi^2$ value here would indicate the possibility
  of ruling out suppression. The value of $\chi^2$ turns out to be 57.8,
  good enough to rule out suppression at roughly 7.6$\sigma$. (Bins without
  vertical error bars contribute nothing to the $\chi^2$ -- in effect,
  their error bars are infinite.) Note the log scale on both axes.}
\label{fig:pkvsk}
\end{center}
\end{figure}

\begin{figure}[t]
\begin{center}
\includegraphics[width=.50\textwidth]{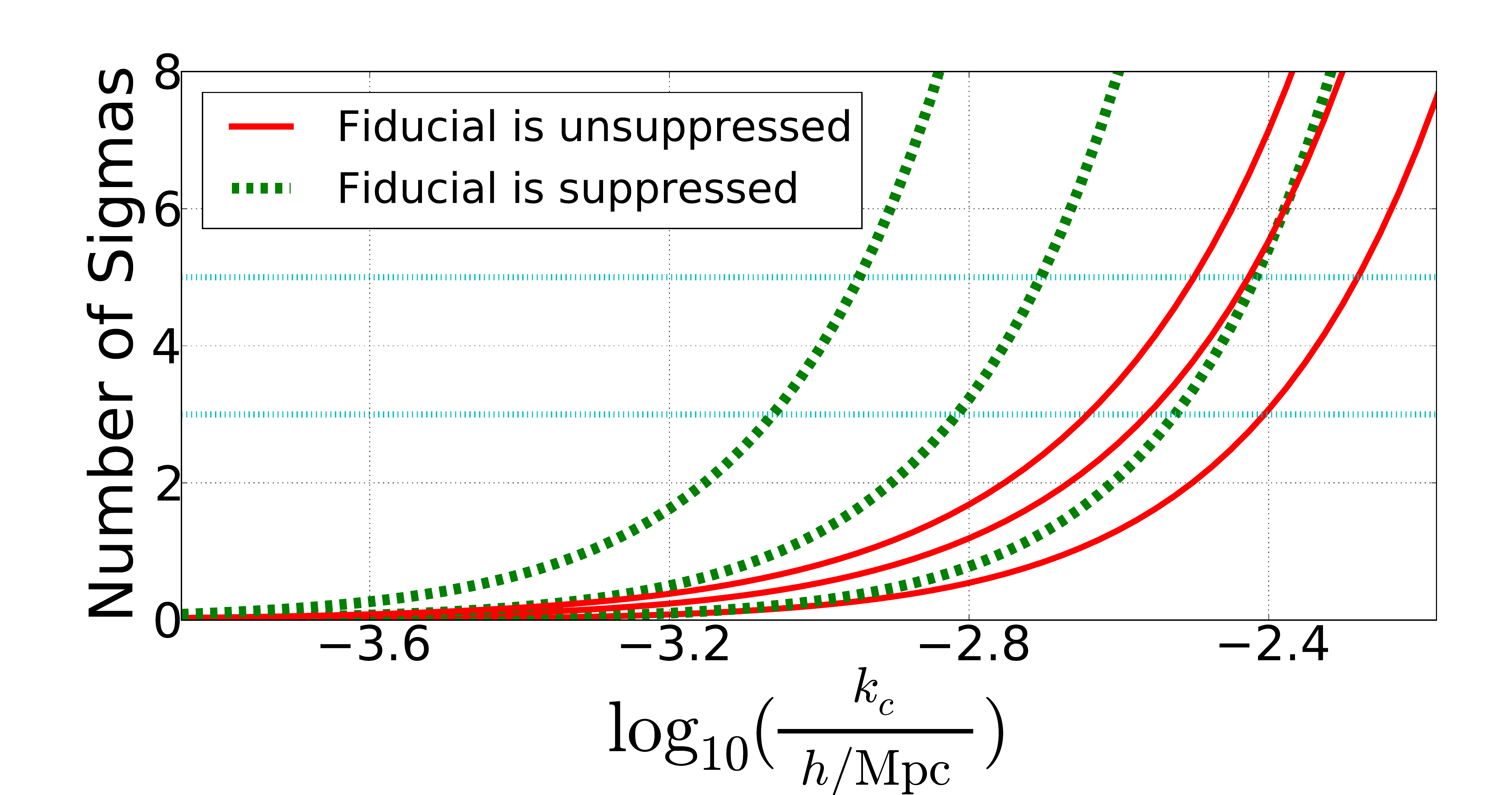}
\caption{The detectability of suppression as a function of
  $k_c$. These results apply to a survey that extends from $z=0$ to
  $z=3$ covering 23,000 square degrees of sky. From bottom to top in
  each set of lines, we assume 0.5, 5 or 50 (spectroscopic) galaxies
  per square arcminute. [Note that the $0.5$ gal/arcmin$^2$ case is
  entirely realistic in the near future, corresponding to the number
  density of spectra planned by e.g.\ BigBoss \cite{BigBoss}, while
  $50$ gal/arcmin$^2$ corresponds to the more aggressive case where
  spectra of most galaxies in a large-volume imaging survey are
  taken.]  Here $\alpha$ is fixed at 3.0 and $\beta$ is fixed at 1.0.}
\label{fig:chisquareresults}
\end{center}
\end{figure}

This is illustrated in Fig.~\ref{fig:pkvsk}. The plot shows the
unsuppressed matter power spectrum, along with the suppressed version
for a particular choice of parameters. The goal of calculating
$\chi^2$ as in Eq.~(\ref{eqn:dchi2}) is to determine whether the
unsuppressed power spectrum can be distinguished from the suppressed
power spectrum for a given set of parameters within the error bars
that would be set by an LSST-like survey. For the case pictured -- in
which the unsuppressed model is taken as true, and the error bars are
calculated based on the suppressed model which is being tested -- it
is possible to rule out suppression with high statistical
significance. The opposite is, however, not true: if the suppressed
model is true, it will be very difficult to rule out the standard
unsuppressed \lcdm due to its larger errors.

For example, survey measurements that fell along the curve predicted
for the unsuppressed $P(k)$ would, in the case shown in
Fig.~\ref{fig:pkvsk}, fall outside some of the suppressed-model error
bars, and ultimately combine to give a total $\chi^2$ of $57.8$, given
the survey parameters outlined in the next paragraph. Meanwhile,
taking the unsuppressed model as fiducial would allow for the
possibility of survey measurements ruling out $\Lambda$CDM, but the
total $\chi^2$ would shrink to $2.0$ due to the larger error bars.

The final results for the detectability of suppression are shown in
Fig.~\ref{fig:chisquareresults}. Instead of plotting $\chi^2$ we show the
number of sigmas (i.e.\ $\sqrt{\Delta\chi^2}$) at which suppressed and
unsuppressed power spectra can be distinguished assuming one degree of freedom
on the measurements of $P(k)$. The figure shows the results as a function of
$k_c$, holding the parameters $\alpha$ and $\beta$ fixed at 3.0 and 1.0,
respectively, and assuming three different possible values for the number of
galaxies observed per square arcminute ($0.5$, $5$, and $50$) in the
spectroscopic survey. We also examined the results with different values of
$\alpha$ and $\beta$, but changes in these parameters do not greatly affect
the results unless $\beta$ becomes close to zero. The scale
of the suppression as determined by $k_c$ is by far the greatest contributing
factor in determining whether a given suppressed model will be detectable to a
large-volume redshift survey.

Comparison of Fig.~\ref{fig:chisquareresults} with Fig.~\ref{fig:pclps12}
shows that if present data for $C_{\ell}$ and $C(\theta)$ truly point to
suppression of the primordial power spectrum, that suppression is likely on
scales that are too large for foreseeable redshift surveys to either detect or
rule out. The most optimistic scenario shown in
Fig.~\ref{fig:chisquareresults}, in which there are 50 galaxies per square
arcminute in the spectroscopic survey, still cannot (at 3$\sigma$) rule out
suppression if $\logkc \lesssim -3.0$,
and cannot rule out \lcdm unless the Universe actually shows suppression of
the matter power spectrum on much smaller scales, with $\logkc \gtrsim -2.7$.
Suppression on scales this small is strongly disfavored by WMAP $C_{\ell}$
observations. Meanwhile, the scales on which WMAP observations tend to favor
suppression ($\logkc \sim -3.3$) are nearly inaccessible to galaxy
surveys. This is a reflection of the fact that the CMB probes much larger
scales than even the largest-volume redshift surveys of the near future.

If {\it only} the cut-sky $S_{1/2}$ statistic is taken into account, CMB
observations heavily favor suppression on scales where suppression would be
readily detectable by redshift surveys, at several sigma, for number densities
of galaxies expected in near-future spectroscopic samples.

\section{Conclusions}\label{sec:conclusions}


In this paper we have studied the suppression of primordial power on large
scales as a possible explanation for the CMB observations. Without considering
particular physical models for the suppression, we adopted a more pragmatic
approach and addressed the following question: do the suppressed models
actually improve the likelihood of the observed CMB sky and, if so, can the
upcoming large-volume galaxy redshift surveys be used to confirm this
suppression?

We first motivated our search by attempting to invert the observations of the
angular power spectrum $C_\ell$ in order to reconstruct the three-dimensional
power spectrum $P(k)$.  As expected, this procedure is very unreliable and
noisy due to the nature of the inverse problem; nevertheless, we obtained
useful hints for the form of the suppression that we should be considering
(see Fig.~\ref{fig:inverseproblemresult} in the Appendix).

We then proceeded to use a parametric model of the suppression
(Eq.~(\ref{eqn:suppressionparameterized})), with the most important
parameter (and the only one we varied in our analysis) being the
suppression scale $k_c$.  We found (see Fig.~\ref{fig:pclps12}) that
the angular power spectrum $C_\ell$, traditionally inferred using
maximum-likelihood-type estimators, prefers a moderate suppression of
power; conversely, the cut-sky pixel-based correlation $C(\theta)$
prefers a stronger suppression. It is also possible that both the
full-sky measurement of $C_\ell$ {\it and} the cut-sky measurement of
$C(\theta)$ are not anomalous, but rather that the underlying
cosmological model is not statistically isotropic. While it is not
clear how to write down the combined likelihood in the full-sky and
cut-sky measurements without assuming statistical isotropy, our simple
choice (Eq.~(\ref{eqn:likelihoods})) prefers the suppression at $\logkc
\approx -3.3$, and increases the combined likelihood by about a factor
of 8.7, corresponding to $\Delta\chi^2=-4.3$.

Detectability of such a large-scale suppression with future surveys will be
difficult, however, as shown in Fig.~\ref{fig:chisquareresults}. In order to
detect the suppression favored by the CMB angular power spectrum, an LSST-type
survey, with a volume of about 100 Gpc$^3$ and a very large number of galaxy
redshifts measured, will be necessary. Roughly speaking, a statistically
significant ruling-out of the power suppression will require spectra taken of
most galaxies in the imaging portion of the survey; this will require a $\sim
10$-meter ground-based, or a $\sim 1.5$-meter space-based, telescope dedicated
to taking spectra. Alternatively, photometric redshift techniques may someday
become so accurate that our preferred case of ``nearly all galaxies being
spectroscopic'' is validated relatively straightforwardly.

Additionally, we point out that suppressed power will be more easily ruled out
(given that the true power is not suppressed) than vice versa, essentially
because the model being tested has smaller cosmic-variance errors if it has
lower power. Therefore, if indeed we live in the universe with the true power
spectrum of density fluctuations being standard inflationary power-law (i.e.\
unsuppressed), then, for example, a survey covering half the sky with 5 galaxy
redshifts per square arcminute will be able to rule out power suppressions on
scales above roughly 1 Gpc at 3$\sigma$ confidence; suppression extending to
smaller scales is even easier to detect. 

Overall, we are optimistic about the prospects of galaxy surveys to
test models of the suppressed large-scale power of primordial
fluctuations. Dark Energy Survey (DES; \cite{des05}), Baryon
Oscillation Spectroscopic Survey (BOSS; \cite{boss}) and, especially,
very-large-volume surveys such as the LSST \cite{LSST}, Joint Dark
Energy Mission (JDEM; \cite{JDEM}), Euclid \cite{Euclid}, and BigBoss
\cite{BigBoss}, will be able to test, at least in part, observations
of CMB experiments on the largest observable scales.

\acknowledgments

We thank Craig Copi, Dominik Schwarz, Glenn Starkman, Roland de Putter,
J\"{o}rg Dietrich, and Andrew Zentner for useful discussions.  The authors are
supported by NSF under contract AST-0807564, NASA under contract NNX09AC89G,
and DOE OJI grant under contract DE-FG02-95ER40899.

\appendix

\section{Direct inversion to obtain the primordial power spectrum}\label{sec:inverse}

As pointed out in Sec.~\ref{sec:prelim}, the CMB angular power spectrum
$C_{\ell}$ is given in terms of the primordial power spectrum by
\begin{eqnarray}
\frac{\ell (\ell + 1) C_{\ell}}{2 \pi} &=& \int d(\ln k) 
\left \lbrack T_{\ell}(k) \right \rbrack ^2 \Delta_{R}^2 (k) \\
& = & \sum_k F_{\ell k} \Delta_R^2(k),
\label{eqn:ClDeltakernelapp}
\end{eqnarray}
where the discretized numerical kernel $F_{\ell k}$ is extracted from CAMB
\cite{CAMB}. Trying to find $\Delta_{R}^2 (k)$ from a given set of
$C_{\ell}$'s (which themselves correspond to a given $C(\theta)$) is an
inverse problem, which we attempted to solve using two different
strategies. [We also attempted a third, doing a simple matrix inversion of the
kernel, but this strategy simply does not work due to the extreme
ill-conditioning.] The first is the Richardson-Lucy method, an algorithm that
iteratively solves for the portion of the sum that multiplies the kernel
\cite{nicholson2009reconstruction,Shafieloo:2003gf,Shafieloo:2007tk,Hamann:2009bz}:

\begin{equation}
  \Delta^2_{i+1}(k) = \Delta^2_i(k) \sum_{\ell = \ell_{\min}}^{\ell_{\max}} 
\tilde F_{\ell k} \frac{C_{\ell}^{obs}}{C_{\ell}^i},
\label{eqn:rleqn}
\end{equation}
where $C_{\ell}^{obs}$ are the observed $C_{\ell}$'s (in this case,
the $C_{\ell}$'s corresponding to the smoothed $C(\theta)$ shown as
the green curve in Fig.~\ref{fig:ctheta}), $C_{\ell}^i$ is calculated
from Eq.~(\ref{eqn:ClDeltakernelapp}) for each iteration $i$, and
\begin{equation}
\tilde F_{\ell k} = F_{\ell k}/\sum_{\ell}F_{\ell k}.
\label{eqn:Ftilde}
\end{equation}
The method converges to a solution for $\Delta_R^2(k)$, but has no
special properties guaranteeing convergence or smoothness of the
solution.

A second strategy makes use of linear regularization, which is one way
of putting extra constraints on the solution. The angular power
spectrum $C_{\ell}$ is a two-dimensional quantity, while the
primordial power spectrum $\Delta^2(k)$ is three-dimensional, and so
finding the latter from the former is an underdetermined
problem. Linear regularization compensates for the fact that the
inverse problem is underdetermined.

Numerical Recipes \cite{NR} outlines one method of regularizing, which
favors a constant solution and penalizes deviations from this. The
goal here is to solve the equation 
\begin{equation}
Ku = b,
\label{eqn:generalform}
\end{equation}
where we refer to matrix $K$ as the kernel; $b$ is a known vector; and $u$ is
the vector to be solved for. Regularization does this by solving the
regularized equation
\begin{equation}
(K^T K + \lambda H)u = K^T b,
\label{eqn:regularizedform}
\end{equation}
where $H$ is a matrix that takes a different form depending on whether the
regularization is linear (penalizing deviations from a constant solution),
quadratic, etc., and $\lambda$ is a parameter that controls how strong the
regularizing constraint is: higher values of $\lambda$ impose stronger
regularizing constraints on the solution. In our case, we take the kernel $K$
to be $F_{\ell k}$, $u$ to be a column vector corresponding to
$\Delta_R^2(k)$, and $b$ to be a column vector corresponding to
$C_{\ell}^{obs}$. We adopt $H$ corresponding to linear regularization.

\begin{figure}[h]
\begin{center}
\includegraphics[width=.45\textwidth]{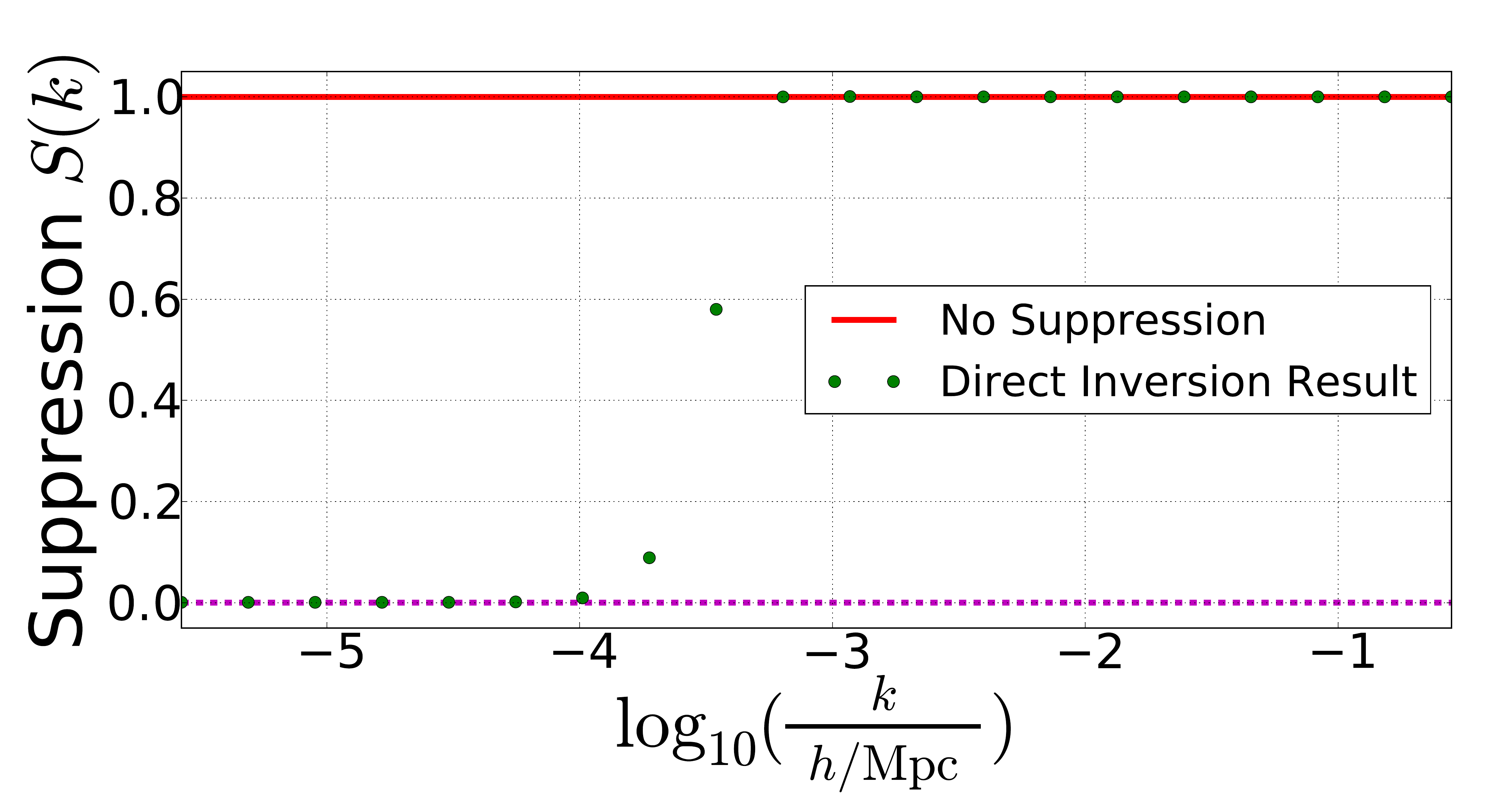}
\caption{Sample results of the direct inversion. The green points
  represent the factor $S(k)$ (see definition of $S(k)$ in
  Eq.~(\ref{eqn:suppressionparameterized})) by which the primordial
  power spectrum $\Delta_R^2$ is suppressed as determined by the
  regularized inversion from Eq.~(\ref{eqn:ClDeltakernelapp}). The
  angular power spectrum $C_{\ell}$ used in this inversion corresponds
  to the $C(\theta)$ shown as the smoothed green curve in
  Fig.~\ref{fig:ctheta}, though we could have in principle used the
  actual WMAP data shown as a blue curve in the same figure.}
\label{fig:inverseproblemresult}
\end{center}
\end{figure}

It is difficult to get consistent results from either of these
strategies, given the ill-conditioned and underdetermined nature of
the inverse problem. This is part of the reason why we chose to focus
most of our attention on doing the forward problem outlined in
Eq.~(\ref{eqn:flowchartbackward}). However, to the extent that
consistent results are possible, both strategies give similar
solutions. A sample result for the suppression factor $S(k)$ (see
definition of $S(k)$ in Eq.~(\ref{eqn:suppressionparameterized})) is
shown in Fig.~\ref{fig:inverseproblemresult}. 

The most notable feature of the inversion result is that it
transitions to near-zero power at large scales/low $k$, with a form
suggesting an exponential cutoff; this provides motivation for
adopting the form we did for parametrizing $\Delta_R^2$ in the forward
problem (Eq.~\ref{eqn:suppressionparameterized})). Unfortunately, a
direct inversion of the sort described in this Appendix requires some
fine-tuning in order to get results of this quality, which is why we
have emphasized that these results are suggestive rather than
conclusive. The results in Fig.~\ref{fig:inverseproblemresult} (which
correspond directly to the regularized inversion, but are also similar
to the results of the Richardson-Lucy method) rely on careful tuning
of the regularization parameter $\lambda$ to ensure that the result
does not become negative at low $k$. [If the regularization is ``not
strong enough,'' with $\lambda$ too small, deviations from a constant
solution are not sufficiently penalized to prevent the solution
becoming negative. If the regularization is ``too strong,'' with
$\lambda$ too large, the solution simply stays constant at roughly
1. Only in particular intermediate cases does it transition nicely
from 0 to 1.]

In addition to this issue, the fact that the inversion results show $S(k)$ of
exactly 1 at high $k$ is a result of a mechanism that was put into the
solution process by hand. Without enforcing the high-$k$ value, the results
often converge to a constant which deviates from unity at high $k$. To
compensate for this and for the issue that the solution does not always
asymptote to nonnegative values at low $k$, we attempted a modification of the
regularized inversion in which deviations from 0 and 1 are penalized (rather
than deviations from a constant solution, as in linear regularization). The
solutions we obtained using this method have the same general form as shown in
Fig.~\ref{fig:inverseproblemresult}, with a transition from 0 to 1 somewhere
between $\logkc = -3$ and $-4$, but the results are even noisier than the
results obtained with Richardson-Lucy and linear regularization.

When the WMAP data was used directly as input, rather than the
$C_{\ell}$ corresponding to the smoothed model in
Fig.~\ref{fig:ctheta}, solutions to the inverse problem were even
noisier. 

Finally, neither the regularized inversion nor the Richardson-Lucy method give
error bars with which the precision of the inversion might be judged. For all
these reasons, the results of the direct inversion cannot be taken as anything
more than suggestive. With that caveat, it is still notable that results of
both the regularized inversion and Richardson-Lucy method \textit{do}
consistently suggest a transition from suppressed power at low $k$ to
unsuppressed power at high $k$. This provides a hint of the fact that the
likelihood of the WMAP $C_{\ell}$ and $C(\theta)$ data may be increased by
introducing suppression, as explored much more fully, and confirmed, in
Sections \ref{sec:statistical} and \ref{sec:constraints}.

\bibliography{cmb_review}

\end{document}